\documentclass[times,a4]{preprint}

\title{A systematic method to enforce conservativity\\ 
on semi-Lagrangian schemes} 

\author{Alexandre \textsc{Cameron}\footnotemark[1] \&
Emmanuel \textsc{Dormy}\footnotemark[2]}
\date{}
\usepackage{MHD}

\begin{document}
\maketitle
\renewcommand{\thefootnote}{\fnsymbol{footnote}}
\footnotetext[1]{
  LPS/LRA,                                                                                                                                
  \'Ecole Normale Sup\'erieure, PSL Research University;                                                                                                                         
  Universit\'e Paris Diderot Sorbonne Paris-Cit\'e;                                                                                                                              
  Sorbonne Universit\'es UPMC Univ Paris 06; CNRS;                                                                                                                               
  24 rue Lhomond, 75005 Paris, France
  (alexandre.cameron@ens.fr)}
  \footnotetext[2]{Department of Mathematics and 
  Applications, CNRS UMR-8553, 
  \'{E}cole Normale Sup\'{e}rieure, 
  45 rue d'Ulm, 75005 Paris, France 
  (emmanuel.dormy@ens.fr)}
\renewcommand{\thefootnote}{\arabic{footnote}}

\section*{Abstract}
Semi-Lagrangian schemes have proven to be very efficient to model 
advection problems. However most semi-Lagrangian schemes are not 
conservative. Here, a systematic method is introduced in order to 
enforce the conservative property on a semi-Lagrangian advection scheme. 
This method is shown to generate conservative schemes with the same 
linear stability range and the same order of accuracy as the initial advection 
scheme from which they are derived. We used a criterion based on the 
column-balance property of the schemes to assess their conservativity 
property. We show that this approach can be used with large CFL numbers 
and third order schemes.


\section{Introduction}
Semi-Lagrangian methods have been demonstrated to be efficient schemes 
to model advection dominated problems. These methods are intensively used
to solve atmospheric and weather problems
\cite{robert_stable_1981,staniforth_semi-lagrangian_1991},
internal geophysics problems \cite{durran_numerical_1999} or plasma simulations 
\cite{besse_semi-lagrangian_2003,crouseilles_conservative_2010,sonnendrucker_semi-lagrangian_1999}.
However, when conservative properties are sought, the method of 
discretisation usually relies on a finite volume discretisation. Conservativity 
is then ensured by canceling fluxes, defined on the computational cell boundaries
\cite{patankar_numerical_1980,eymard_finite_2000}.

Semi-Lagrangian methods, on the contrary, are in general not conservative. 
Some earlier work have tried to address this issue and derive a 
conservative semi-Lagrangian scheme. For example, 
\cite{nakamura_exactly_2001,xiao_completely_2001} introduced a modified version
of a non-conservative semi-Lagrangian scheme \cite{takewaki_cubic_1985} 
to enforce conservativity. Their approach provides a conservative formulation 
at the cost of introducing a scheme in which the coefficients depends on 
the values of the advected field. An alternative strategy which uses a 
semi-Lagrangian reconstruction to estimate fluxes on the faces, was introduced by 
\cite{crouseilles_conservative_2010} in the finite volume spirit to model the 
Vlasov equation. This strategy was adapted to compressible flow in 
\cite{qiu_conservative_2011}.
In both of the above approaches, the formulations are well adapted
to one-dimensional problems, their generalizations to higher spatial dimensions
without using a splitting strategy is challenging.

A general method to enforce conservativity on a semi-Lagrangian scheme
was introduced in Lentine {\it et al.} \cite{lentine_unconditionally_2011}. 
Noting that the contribution of a given cell to the update 
in time of the total field does not add up to unity, they introduced an {\it ad hoc} 
modification of the coefficients which allows to ensure conservativity at the cost of 
reducing the order of the scheme. 

We propose a systematic method to enforce conservativity on a numerical scheme.
Our method, follows ideas introduced in the method of support operators developed
by Shashkov \cite{shashkov_conservative_1995}, or the summation by part
method of Carpenter {\it et al.} \cite{carpenter_stable_1999,fisher_discretely_2013}. 
It can easily be applied to semi-Lagrangian schemes. A close equivalence can be 
found with the flux interpretation in the sense of finite volume schemes.
Let us start by considering the continuity equation, for a quantity $\Phi$ 
subject to a velocity field $\U$
\begin{align}
	\partial_t \Phi = - \diV ( \Phi \U ) \equiv \Ctn [\Phi] \, ,
	\label{eq:CSV}
\end{align}
where \Ctn{} denotes the continuity operator.
If the flow is incompressible, $\diV \U =0$, the continuity equation
reduces to the advection equation:
\begin{align}
	\partial_t \Phi = - \Phi (\diV \U) - ( \U \cdot \grad ) \Phi 
	= - (\U \cdot \grad ) \Phi \equiv \Dtn [\Phi] \, ,
	\label{eq:ADV}
\end{align}
where \Dtn{} denotes the advection operator. 

Instead of considering eq.~\eqref{eq:ADV} as a simplified version of 
eq.~\eqref{eq:CSV}, under the solenoidal constraint, the
two equations can be viewed as two independent equations. 
Introducing the canonical scalar product of two continuous fields 
$\Psi$ and $\Phi$,
$ \scalpr{ \Psi }{ \Phi } = \int \Psi \Phi \ddif \tau \, , $
the continuity and advection operators are then adjoint operator 
up to a change of the velocity sign:
\begin{align}
	\int \Psi \left [ (\U \cdot \grad) \Phi \right ] \ddif \tau
		 = \oint \Psi \Phi \U \cdot \nvec \ddif s + 
	\int \left[ \diV \left ( - \U \Psi \right ) \right]\Phi \ddif \tau\, .
	\label{eq:adjoint}
\end{align}
If the boundary term vanishes, the operators follow:
$ \scalpr{ \Psi }{ \Dtn \Phi} = \scalpr{ \CtnM \Psi }{ \Phi } $. Such 
relations have been intensively used in the support operator formalism
\cite{shashkov_conservative_1995}. Introducing the $\star$ to denote 
the adjoint operator, we get: $	\Dtn ^{\star} = \CtnM \, .$ This adjoint 
property can be used to enforce conservativity on an arbitrary advection scheme.

\section{Column-balance criterion \& Adjoint operator}
\label{sec:colbal}
Using a linear finite difference scheme explicit on time,
the advection equation is given by:
$
	\Phi^{n+1}_i = \Phi^n_i + \dij \Phi^n_{j} \, ,
$,
where $D_{i,j}$ denotes the discrete linear operator associated to 
eq.~\eqref{eq:ADV}. For the discrete operator to be homogeneous, 
the coefficients $D_{i,j}$ must only depend on the reduced velocity 
$\uci = u_i \Delta t /\Delta x $. 

In a similar way, finite difference schemes modeling the continuity 
equation eq.~\eqref{eq:CSV} can be written as 
$ \Phi^{n+1}_{i} = \Phi^{n}_i + \cij \Phi^{n}_j $
where \cij{} denotes the conservative transport matrix.
The evolution of the total mass, $M$, is then given by 
$
M^{n+1}-M^{n}=\sum_{i,j} \cij \Phi^{n}_j =
\sum_{i} \Phi^{n}_i (\sum_{j} \transp{\cij}) \, .
$
It follows that the scheme is conservative if and only if the \cij{} operator 
is column-balanced, \textit{i.e.} for all $i$, $\sum_j C_{j,i} =0 \, .$
In order to link this formalism to finite volume schemes, the
column-balanced conservative matrix can be compared to the 
flux method. On a regular Cartesian grid, flux are defined
at the boundary between two vertices. 
The equation modeling the flux methods is:
$
	\Phi^{n+1}_i = \Phi^n_i + F_{i-1/2} - F_{i + 1/2} \, , 
$,
where $\Phi^{n+1}_i$ denotes the values of field $\Phi$ at the index $i$ 
and $ F_{i + 1/2} $ the flux of field $\Phi$ computed at index $i + 1/2$. 
Choosing $F_{i + 1/2} = C_{i+1,i} \Phi_i - C_{i,i+1} \Phi_{i + 1} $,
both methods are strictly equivalent.

The adjoint relation will now be used to show how a generic advection 
scheme can be modified to enforce the conservativity property. 
Once the problem is discrete, the adjoint property leads to:
$\cu = \transp{\duM}$. It is a property of the 
transpose that \cu{} and \duM{} have the same eigenvalues.
Both operators are thus stable for the same set of parameters.
They also imply that the error of the \cu{} scheme 
is the transpose of the error of the \duM{} scheme, therefore the two 
operators have the same consistency order. In addition, if \duM{} is 
monotone, \cu{} is also monotone. Using the Lax-Richtmyer 
equivalence theorem \cite{lax_survey_1956}, the consistent and stable 
\cu{} scheme converges to the continuity equation.

The above remarks do not ensure that the \cu{} scheme conserves the 
total mass. However, assuming that the advective scheme is strictly 
consistent, \textit{i.e.} $\forall i \,, \; \sum_j \dijuM = 0$, it follows that
$\forall j, \quad \sum_i \ciju= 0 $. The \ciju{} operator is 
thus column-balance and conserves the total mass.

It is important to stress that we only introduce a modification of the spatial 
operator. The conservative property of \ciju{} will thus be valid both for 
multi-step and multi-stage time-stepping. Consider for example, a Crank-Nicholson
time-stepping scheme \cite{crank_practical_1947,durran_numerical_1999}, the fields 
at each time steps are related via
\begin{align}
	\left( \delta - \frac{\Delta t }{2} D(U^{n+1}) \right)_{i,j} \Phi^{n+1}_j &= 
	\left( \delta + \frac{\Delta t }{2} D(U^{n}) \right)_{i,j} \Phi^{n}_j \,,
	\label{eq:CNformula}
\end{align}
where $\delta_{i,j}$ denotes the Kronecker delta ($\delta_{i,j}=1$ if
$i=j$ and $\delta_{i,j}=0$ if $i \neq j$).
The Crank-Nicholson advection operator ($CN$) can be rewritten
\begin{align}
	CN_{i,j} (U) &= -\delta_{i,j} + \left[\left( \delta- \frac{\Delta t }{2} D(U^{n+1}) \right) ^{-1}
	\left( \delta + \frac{\Delta t }{2} D(U^{n}) \right) \right]_{i,j}
	\, .
	\label{eq:CNeq}
\end{align}
The corresponding conservative operator ($CCN$) is then
\begin{align}
	CCN_{i,j} (U)
	&= 
	-\delta_{i,j} + \left[
	\left( \delta + \frac{\Delta t }{2} \transp{D}(-U^{n}) \right) 
	\left( \delta - \frac{\Delta t }{2} \transp{D}(-U^{n+1}) \right) ^{-1}
	\right]_{i,j}
	\, .
	\label{eq:CCNeq}
\end{align}

\section{Conservative semi-Lagrangian scheme in one dimension}
\label{sec:consSL}
Let us now turn to semi-Lagrangian schemes. The conservative method 
can be used to generate conservative scheme from a semi-Lagrangian 
algorithm. In one dimension, the \CIR{} scheme \cite{courant_solution_1952}, 
which is equivalent to the upwind scheme, will be used to show how a 
conservative \CIR{} (\CCIR{}) scheme can be built. The resulting \CCIR{} 
scheme will then be tested on a simple numerical simulation.

In order to be stable, advection algorithm must transport information 
in the direction of the flow. The \CIR{} scheme satisfies this condition by 
adapting its stencil according to the direction of the velocity following the 
characteristic. For advection equation in one dimension 
(e.g.~\cite{toro_riemann_2009}), the \CIR{} scheme is:
\begin{align}
\label{eq:Nlin}
	&\Phi^{n+1}_i = \Phi^n_i + \CIR_{i,j} \Phi^n_j
	= \Phi^n_i + ( \UU^+_i\Phi_{i-1} - \vert \UU_i\vert \Phi^n_{i} -\UU^-_i\Phi^n_{i+1}) \, ,
\end{align}
with $\UU^{+}_i = \max ( \UU_i ,\, 0 )$ and $\UU^{-}_i = \min ( \UU_i ,\, 0 ).$
To leading order, this scheme yields the diffusive error term
\begin{align}
	\sme{[\partial_t \Phi + u \partial_x (\Phi)]}{\CIR}
	\simeq 
	\frac{\Delta x}{2} \, \vert u \vert \partial_x \left[ 
	\left (1- \ua \right) 
	\partial_x \Phi \right] 
	\, .
	\label{eq:modCIR}
\end{align}
The scheme is consistent with the advection equation, but it is not conservative.
The conservative counterpart of the \CIR{} scheme, can be built by changing the 
sign of the velocity and transposing the $\CIR_{i,j}$ matrix. 
The expression of the \CCIR{} scheme is:
\begin{align}
	\Phi^{n+1}_i = \Phi^n_i + \CCIR_{i,j} \Phi^n_j
	= \Phi^n_i + ( \UU^+_{i-1}\Phi^n_{i-1} - \vert \UU_i\vert \Phi^n_{i} 
	- \UU^-_{i-1}\Phi^n_{i+1})
	\,.
	\label{eq:Nmat}
\end{align}
The \CCIR{} scheme is conservative because it is column-balanced
by construction. Similarly to the \CIR{} scheme, the \CCIR{} scheme 
has a diffusive error. As expected, the \CCIR{} error term is the adjoint 
of the \CIR{} error term:
\begin{align}
	\sme{\left[ \partial_t \Phi + \partial_x (u\Phi) \right] }{\CCIR}
	\simeq 
	\frac{\Delta x}{2}\, \partial_x \left[ 
	\left(1-\ua \right) 
	\partial_x (\vert u \vert \Phi) \right] 
	\, . 
	\label{eq:modMSL}
\end{align}

The \CCIR{} scheme was tested using a velocity profile $ u(t; x) = \sin( 2\pi x ) $ 
and a uniform passive scalar $\Phi (t=0;x) = 1$. It conserved the total mass, 
$M/M_0$, near unity up to machine precision. This is not the case of the \CIR{} 
scheme for varying velocities:
\begin{align}
	\Phi^{n+1}_i = \Phi^n_i + \tfrac{\UU_{i-1}}{\UU_i} (\UU^+_i \Phi^n_{i-1}) 
	- ( \vert \UU_i \vert \Phi^n_{i-1}) 
	+ \tfrac{\UU_{i+1}}{\UU_i} ( \UU^-_i \Phi^n_{i-1}) \,.
\end{align}

In the same manner, the second order (dispersive) Lax-Wendroff scheme (\LW), 
which takes the form:
\begin{align}
	\Phi^{n+1}_i =&
	\left(\tfrac{\UU^{+}}{\UU}\tfrac{\UU(1+\UU)}{2}\right)_{i} \Phi^{n}_{i-1} + 
	\left(\tfrac{\UU^{+}}{\UU}(1-\UU^2)\right)_{i} \Phi^{n}_{i} - 
	\left(\tfrac{\UU^{+}}{\UU}\tfrac{\UU(1-\UU)}{2}\right)_{i} \Phi^{n}_{i+1}
	\\& \nonumber +
	\left(\tfrac{\UU^{-}}{\UU}\tfrac{\UU(1+\UU)}{2} \right)_{i} \Phi^{n}_{i+1} + 
	\left(\tfrac{\UU^{-}}{\UU}(1-\UU^2)\right)_{i} \Phi^{n}_{i} - 
	\left(\tfrac{\UU^{-}}{\UU}\tfrac{\UU(1-\UU)}{2}\right)_{i} \Phi^{n}_{i-1}
	\,,
\end{align}
can be transformed to a conservative \LW{} scheme (\CLW), of the form,
\begin{align}
	\Phi^{n+1}_i =&
	\left(\tfrac{\UU^{+}}{\UU}\tfrac{\UU(1+\UU)}{2}\Phi^{n}\right)_{i-1} + 
	\left(\tfrac{\UU^{+}}{\UU}(1-\UU^2) \Phi^{n}\right)_{i} - 
	\left(\tfrac{\UU^{+}}{\UU}\tfrac{\UU(1-\UU)}{2} \Phi^{n}\right)_{i+1}
	\\& \nonumber + 
	\left(\tfrac{\UU^{-}}{\UU}\tfrac{\UU(1+\UU)}{2} \Phi^{n} \right)_{i+1} + 
	\left(\tfrac{\UU^{-}}{\UU}(1-\UU^2) \Phi^{n}\right)_{i} - 
	\left(\tfrac{\UU^{-}}{\UU}\tfrac{\UU(1-\UU)}{2}\Phi^{n}\right)_{i-1}
	\,.
\end{align}
In the same way, the third order (hyperdiffusive) semi-Lagrangian 
Dahlquist and Björck scheme (\DB) 
(e.g. \cite{durran_numerical_1999,dahlquist_numerical_1974}):
\begin{align}
	\Phi^{n+1}_i =& -\left(\tfrac{\UU^{+}}{\UU}\tfrac{\UU(1-\UU^2)}{6}\right)_{i} \Phi^{n}_{i-2} + 
	\left(\tfrac{\UU^{+}}{\UU}\tfrac{\UU(1+\UU)(2-\UU)}{2}\right)_{i} \Phi^{n}_{i-1}
	\\& \nonumber 
	+\left(\tfrac{\UU^{+}}{\UU}\tfrac{(1-\UU^2)(2-\UU)}{2}\right)_{i} \Phi^{n}_{i} - 
	\left(\tfrac{\UU^{+}}{\UU}\tfrac{\UU(1-\UU)(2-\UU)}{6}\right)_{i} \Phi^{n}_{i+1}
	\\& \nonumber 
	-\left(\tfrac{\UU^{-}}{\UU}\tfrac{\UU(1-\UU^2)}{6}\right)_{i} \Phi^{n}_{i+2} + 
	\left(\tfrac{\UU^{-}}{\UU}\tfrac{\UU(1+\UU)(2-\UU)}{2} \right)_{i} \Phi^{n}_{i+1} 
	\\& \nonumber 
	+ \left(\tfrac{\UU^{-}}{\UU}\tfrac{(1-\UU^2)(2-\UU)}{2}\right)_{i} \Phi^{n}_{i} - 
	\left(\tfrac{\UU^{-}}{\UU}\tfrac{\UU(1-\UU)(2-\UU)}{6}\right)_{i} \Phi^{n}_{i-1}
	\,,
\end{align}
has the following conservative counterpart (\CDB):
\begin{align}
	\Phi^{n+1}_i =& -\left(\tfrac{\UU^{+}}{\UU}\tfrac{\UU(1-\UU^2)}{6} \Phi^{n}\right)_{i-2} + 
	\left(\tfrac{\UU^{+}}{\UU}\tfrac{\UU(1+\UU)(2-\UU)}{2}\Phi^{n}\right)_{i-1} 
		\\& \nonumber 
	+ \left(\tfrac{\UU^{+}}{\UU}\tfrac{(1-\UU^2)(2-\UU)}{2} \Phi^{n}\right)_{i} - 
	\left(\tfrac{\UU^{+}}{\UU}\tfrac{\UU(1-\UU)(2-\UU)}{6}\Phi^{n}\right)_{i+1}
		\\& \nonumber 
	-\left(\tfrac{\UU^{-}}{\UU}\tfrac{\UU(1-\UU^2)}{6} \Phi^{n}\right)_{i+2} + 
	\left(\tfrac{\UU^{-}}{\UU}\tfrac{\UU(1+\UU)(2-\UU)}{2}\Phi^{n}\right)_{i+1} 
		\\& \nonumber 
	+ \left(\tfrac{\UU^{-}}{\UU}\tfrac{(1-\UU^2)(2-\UU)}{2} \Phi^{n}\right)_{i} - 
	\left(\tfrac{\UU^{-}}{\UU}\tfrac{\UU(1-\UU)(2-\UU)}{6}\Phi^{n}\right)_{i-1}
	\,.
\end{align}

\begin{figure}
	\centerline{
	\includegraphics[width=0.5\textwidth] {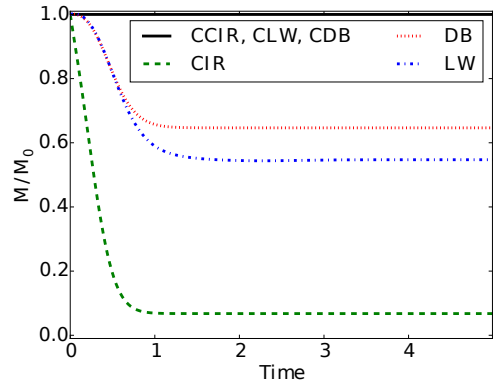}
	} \caption{Comparison of the total mass evolution for the $1$D transport 
	problem with $u=\sin(x)$ and $\Phi(t=0)=1$ 
	for conservative and non-conservative 
	semi-Lagrangian schemes of various order (the CFL number is here 
	fixed to $0.75$).}
	\label{fig:mass1D}
\end{figure}

\begin{figure}
	\centerline{
	\ (a)\includegraphics[width=\tlwidth, trim= 2 2 55 44, clip=true] {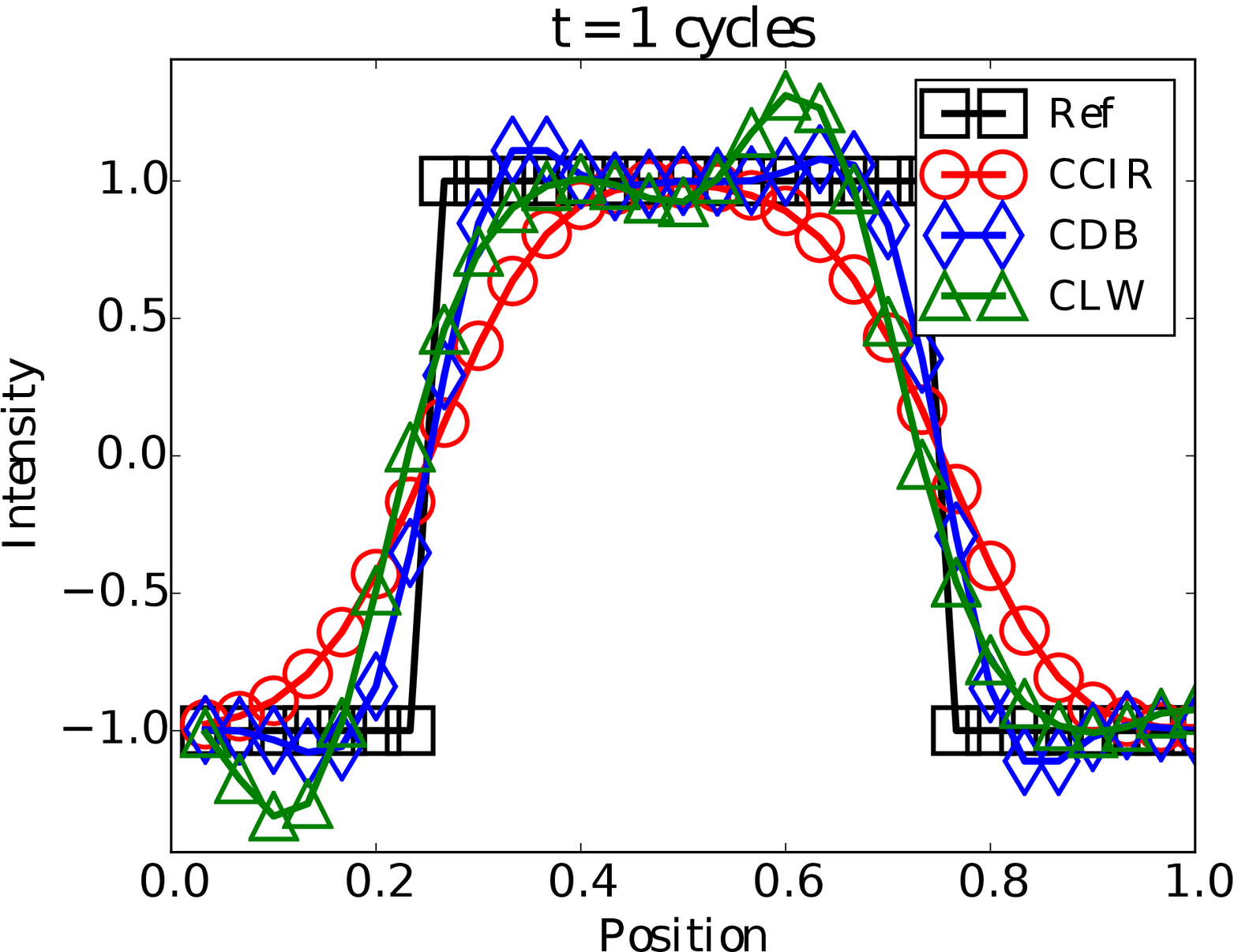}
	\includegraphics[width=\twidth, trim= 20 2 55 44, clip=true] {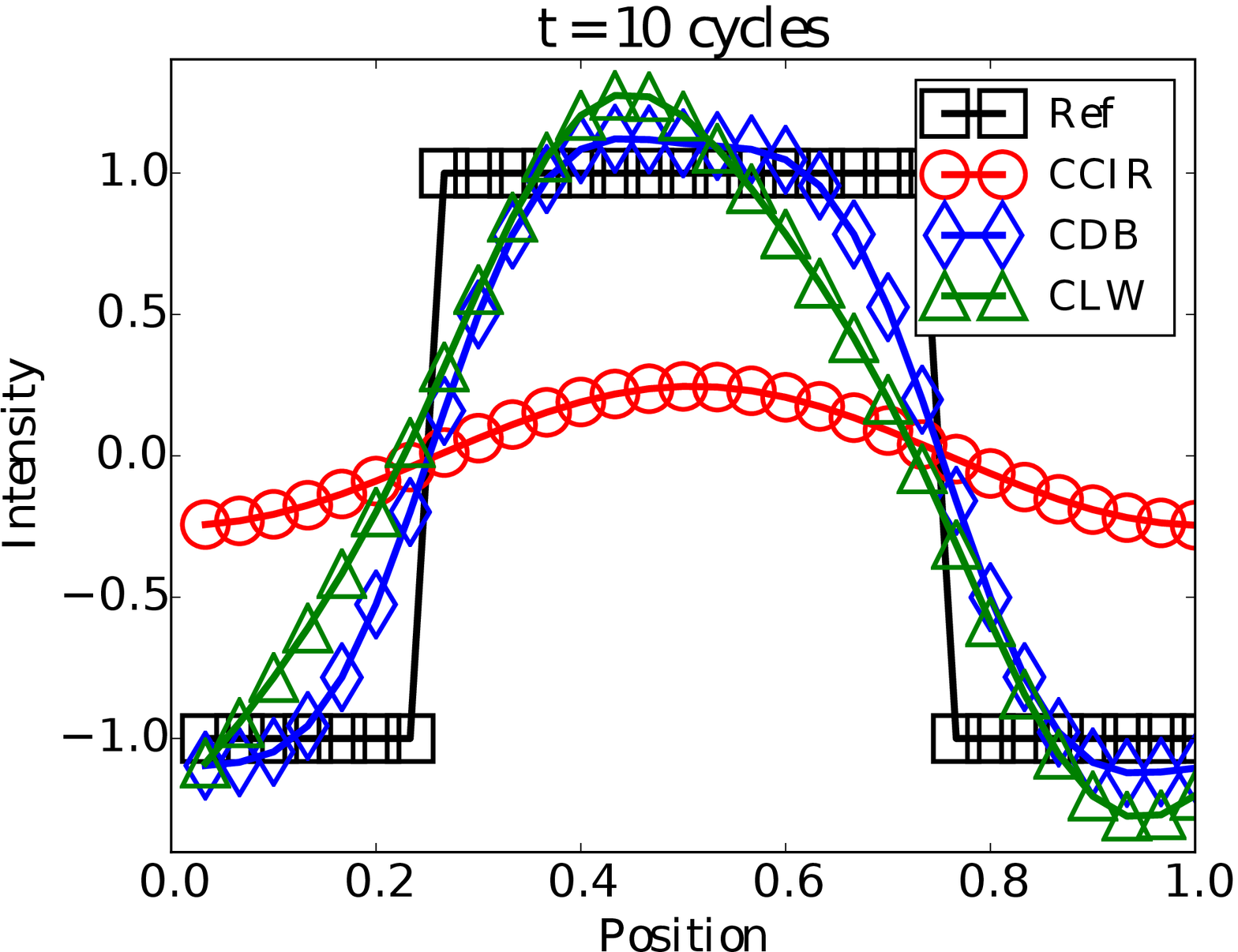}
	\includegraphics[width=\twidth, trim= 20 2 55 44, clip=true] {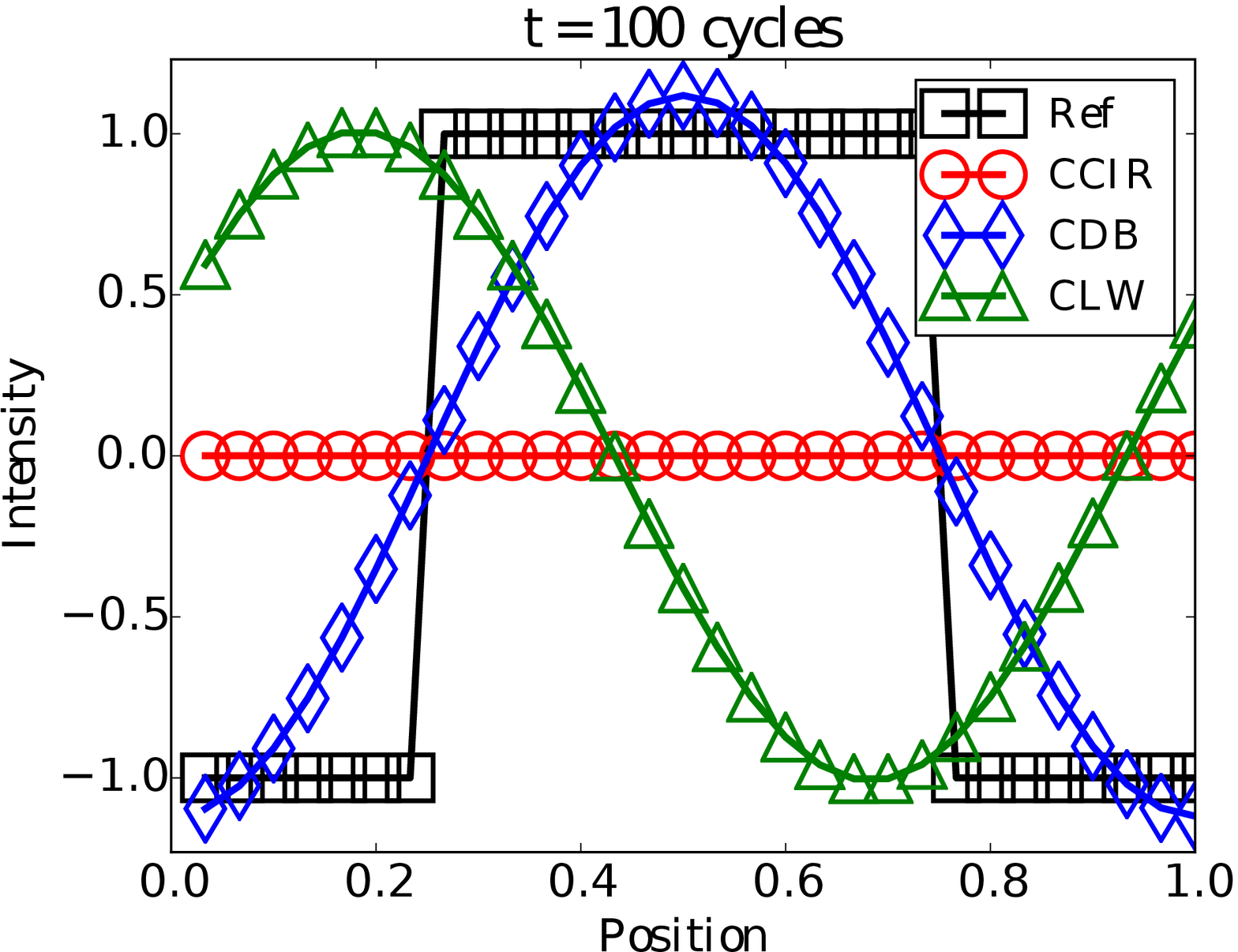}
	}
	\centerline{
	\ (b)\includegraphics[width=\tlwidth, trim= 2 2 55 43, clip=true] {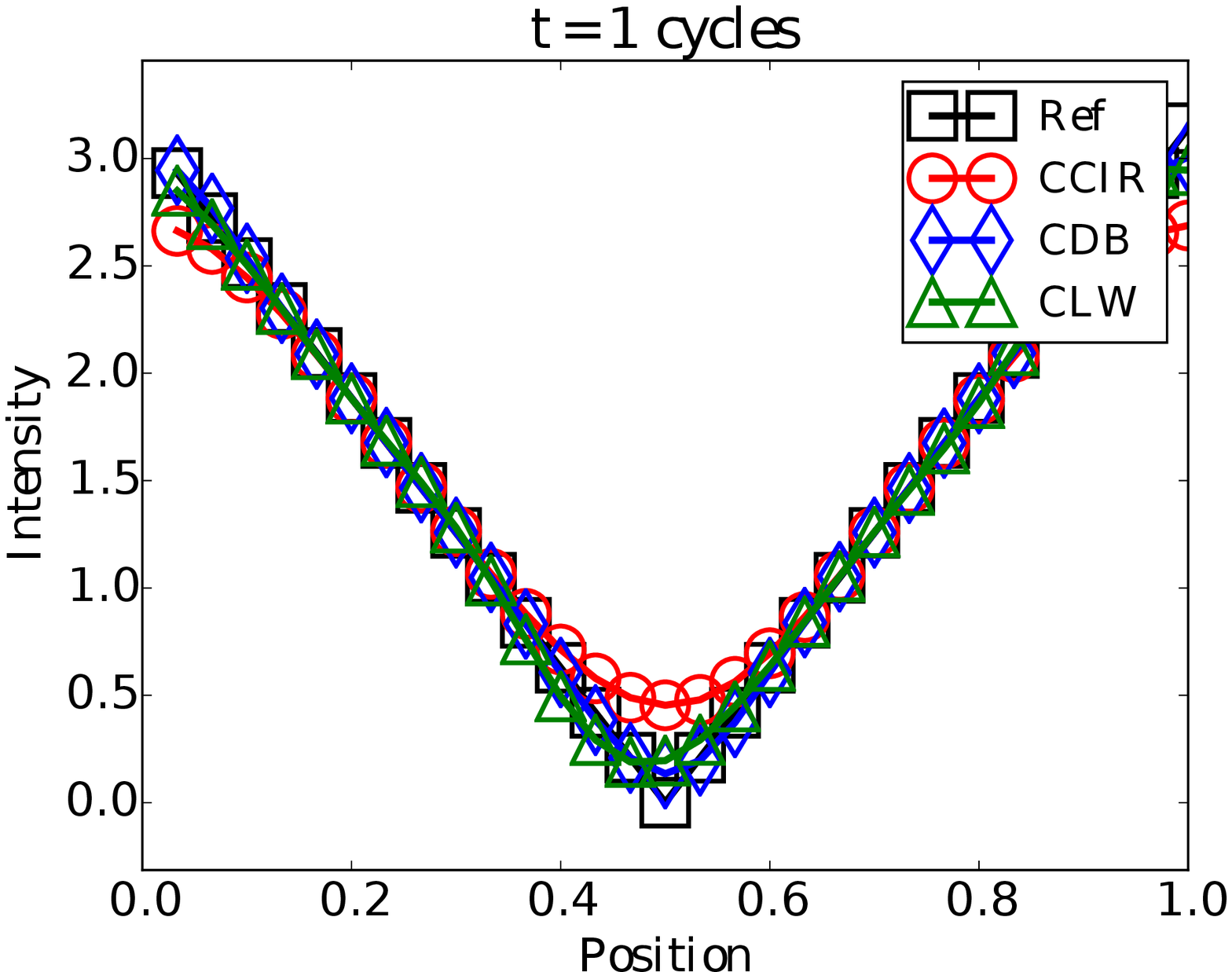}
	\includegraphics[width=\twidth, trim= 20 2 55 43, clip=true] {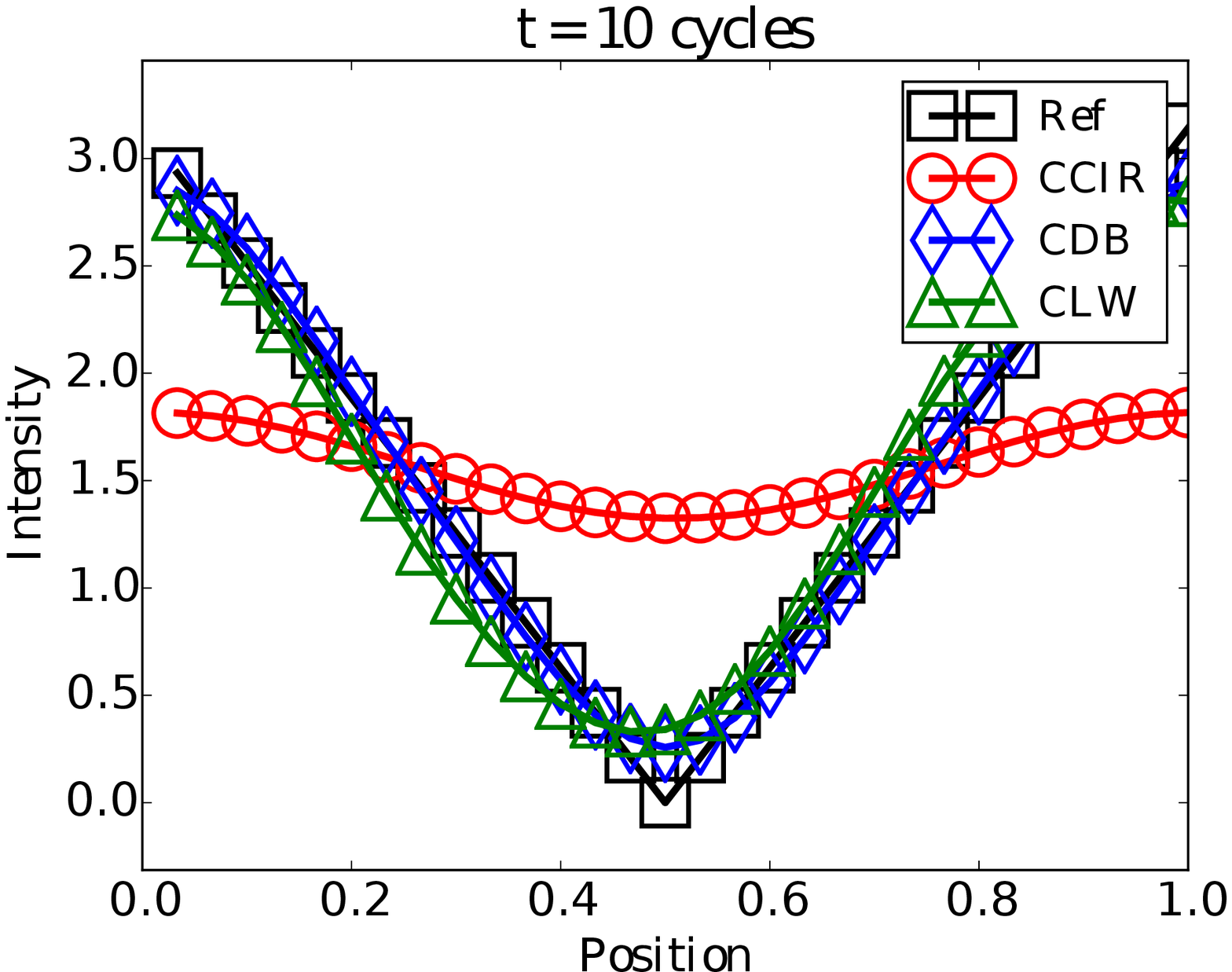}
	\includegraphics[width=\twidth, trim= 20 2 55 43, clip=true] {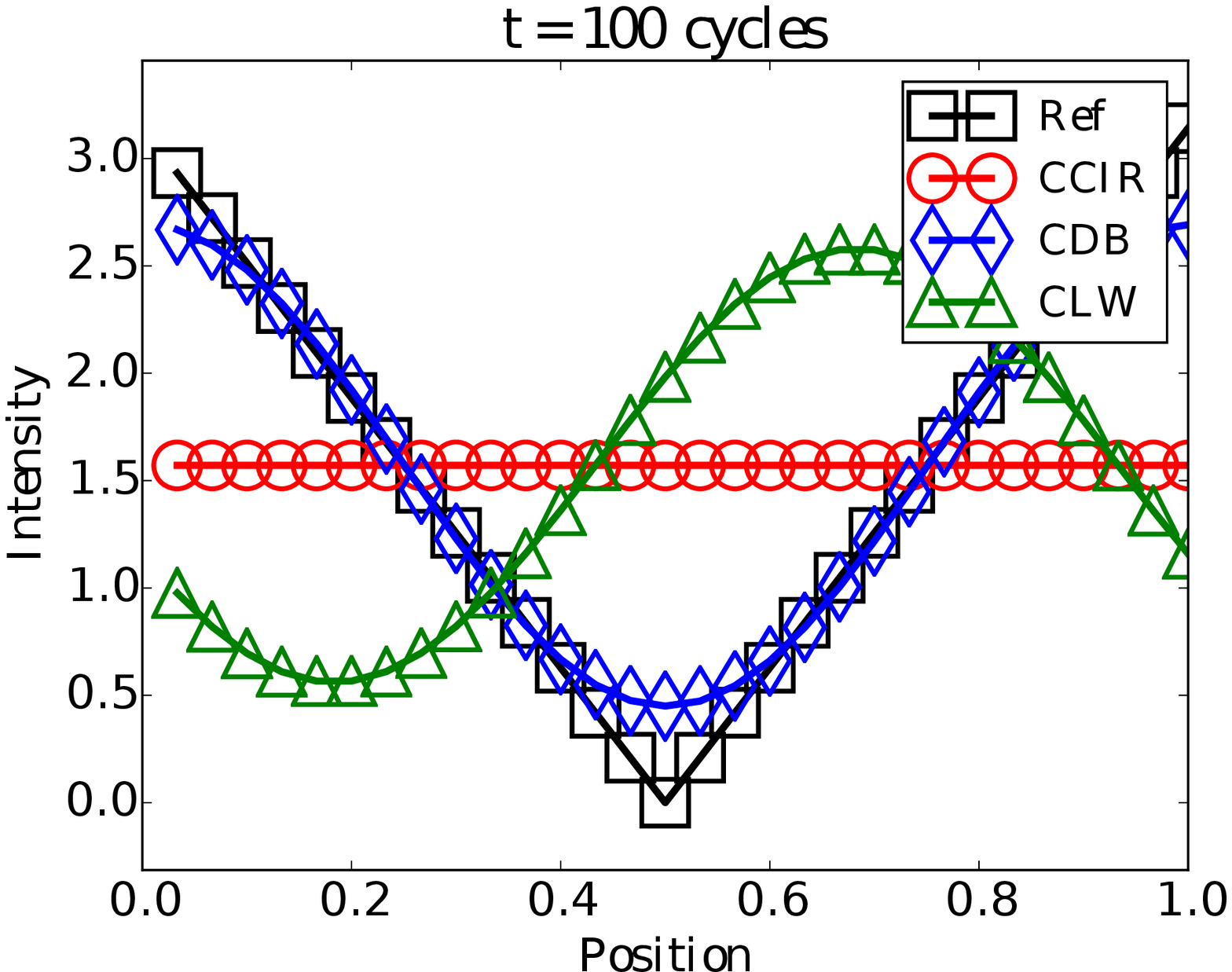}
	}
	\centerline{
	\ (c)\includegraphics[width=\tlwidth, trim= 2 2 55 43, clip=true] {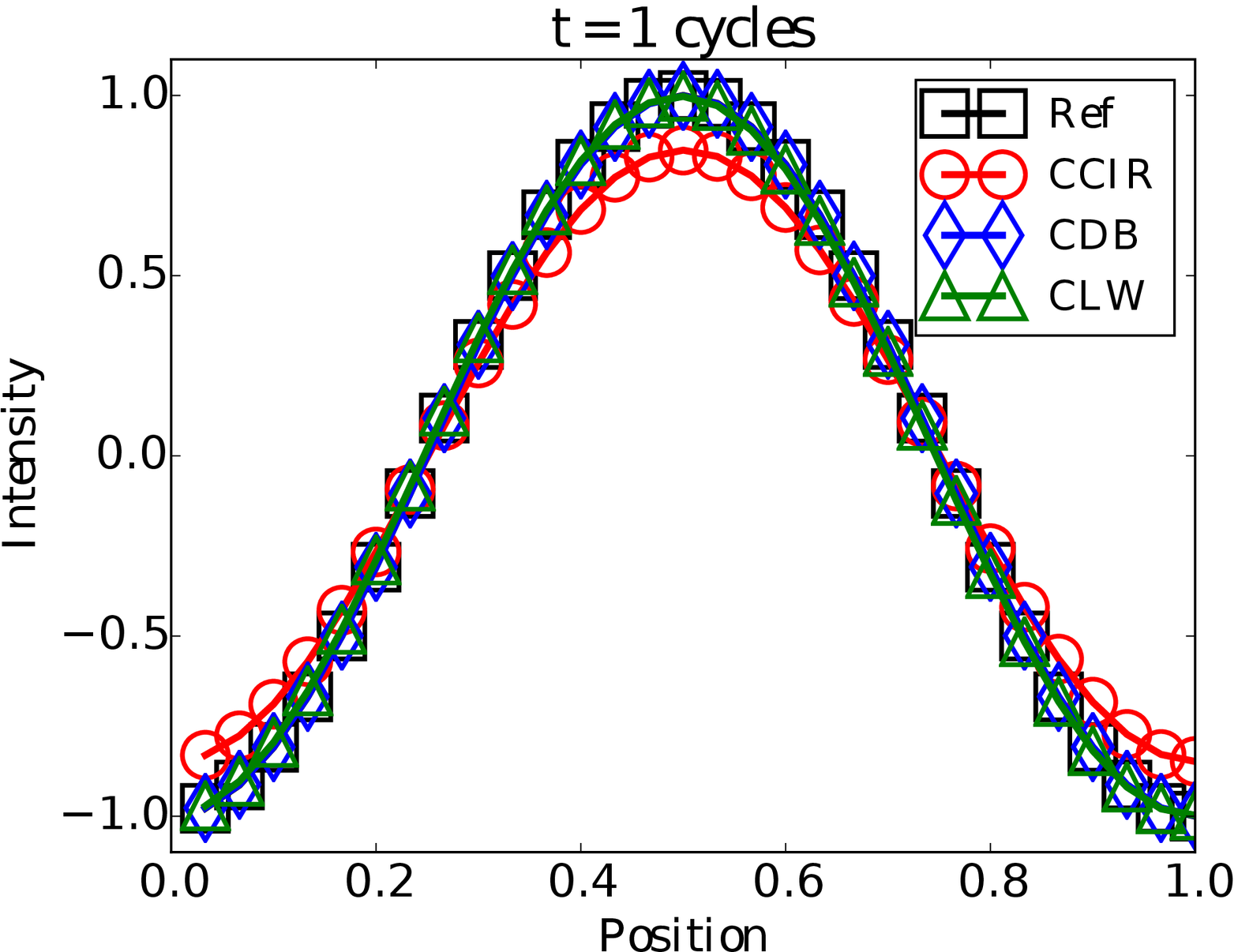}
	\includegraphics[width=\twidth, trim= 20 2 55 43, clip=true] {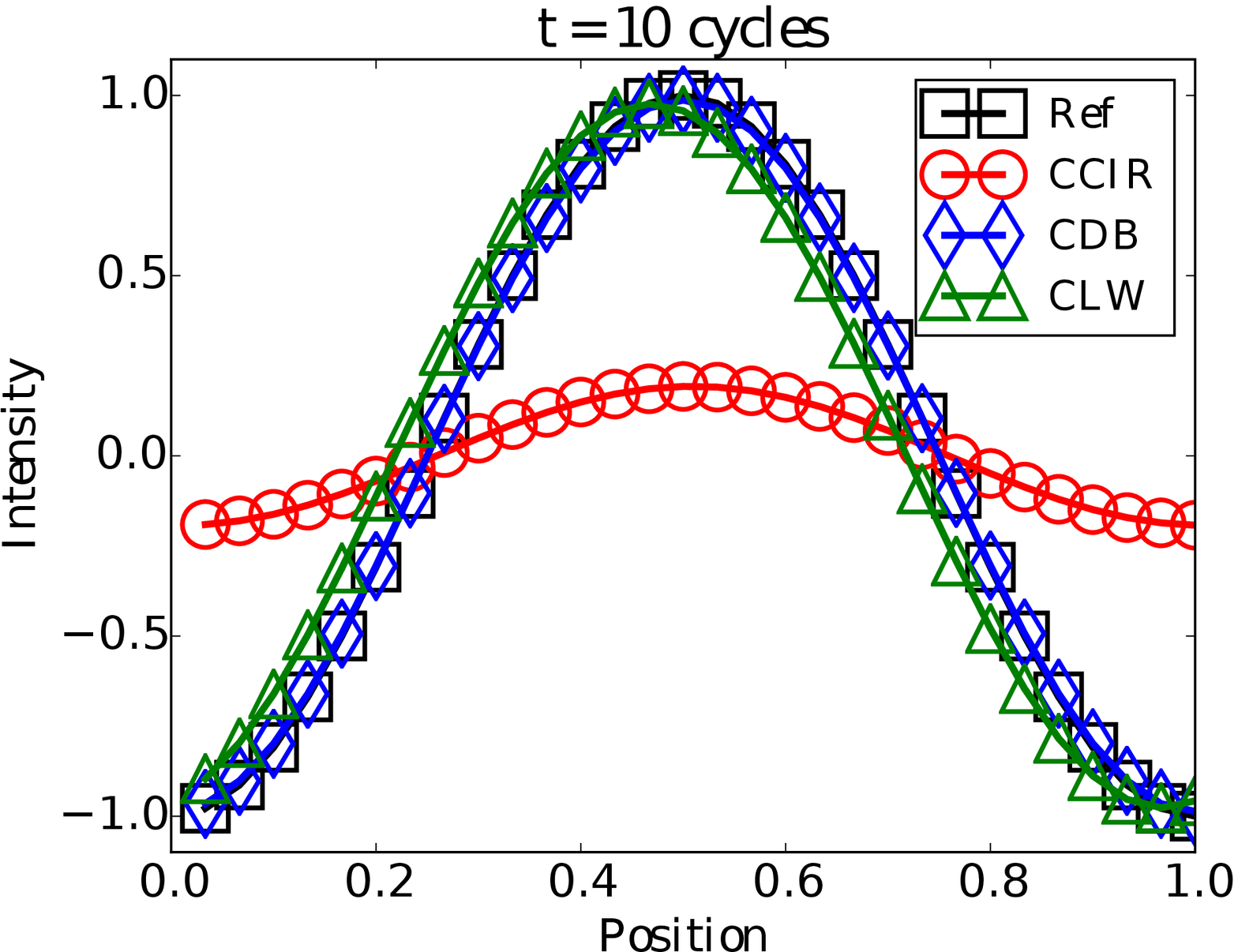}
	\includegraphics[width=\twidth, trim= 20 2 55 43, clip=true] {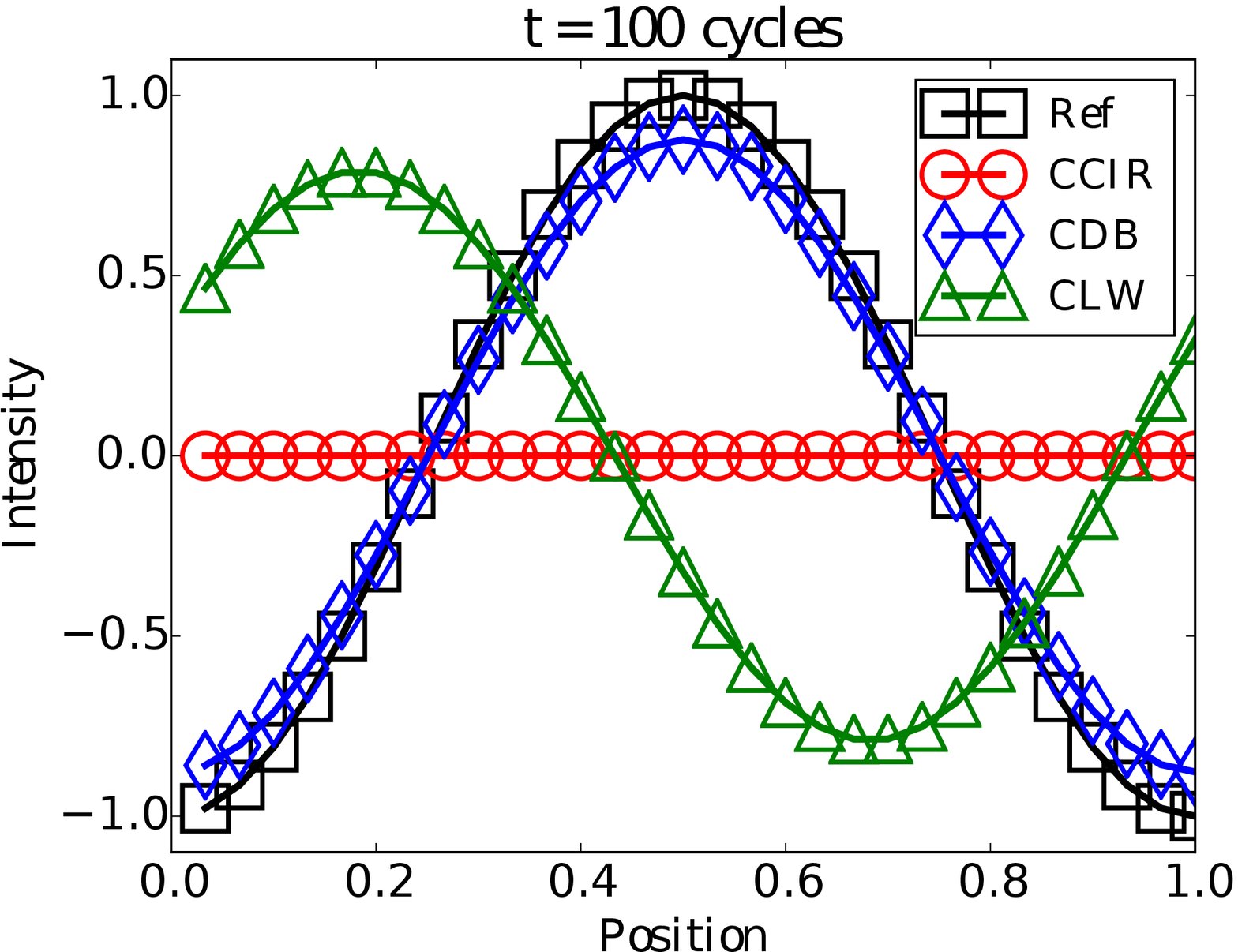}
	}
	\caption{Advection in a periodic domain with periodic boundary
	conditions. The advection velocity is constant and the initial 
	profile takes the form of a Heaviside (a), a piecewise afine (b), 
	a cosine (c) function. Graphes from left to right correspond to 
	$1$, $10$ and $100$ periods of the flow respectively.}
\label{fig:testHTC}
\end{figure}

\begin{figure}
	\centerline{
	\subfigure[Growth-rate]{
	\includegraphics[width=0.4\textwidth, trim= 2 2 45 30, clip=true] {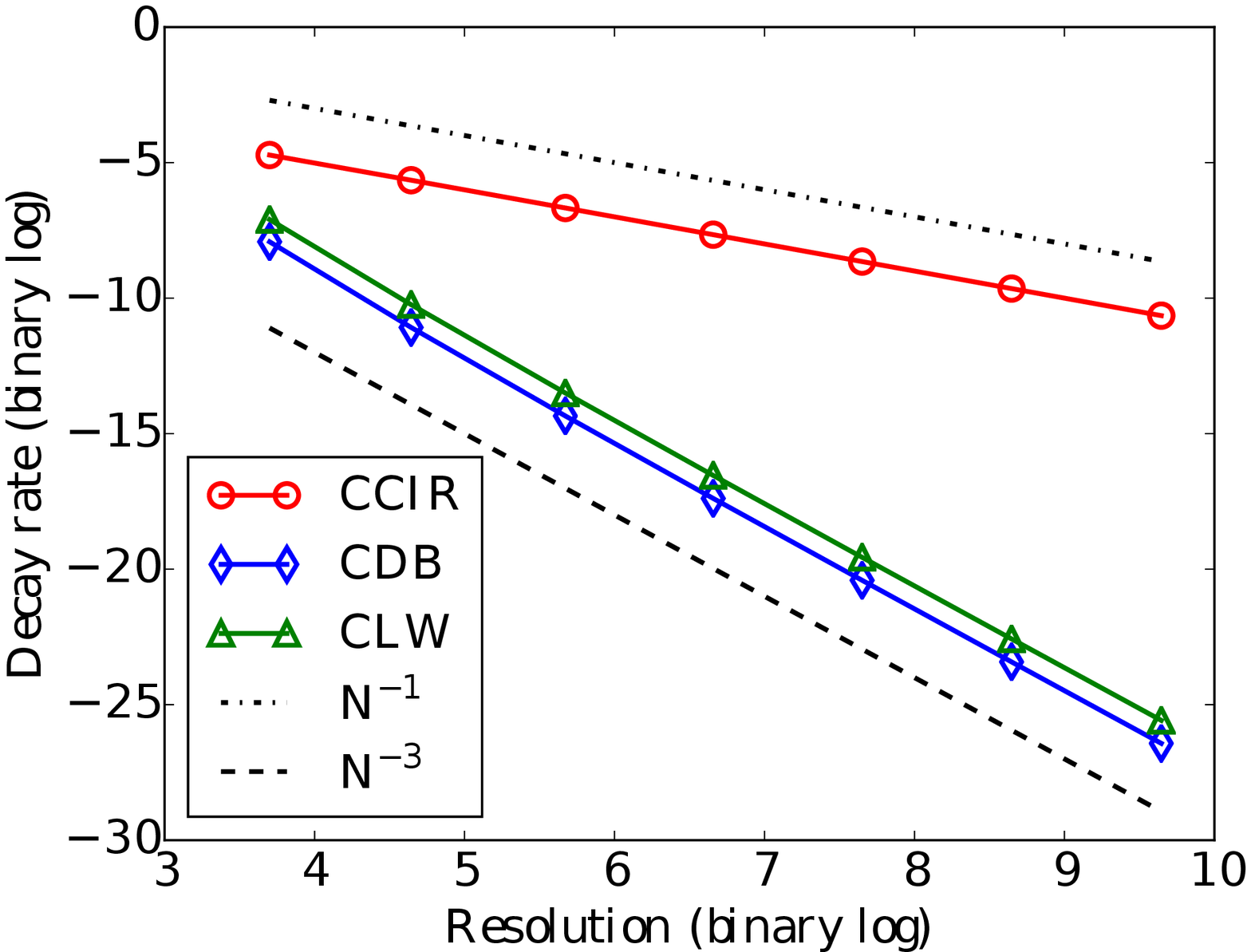}
	}
	\subfigure[Phase-drift]{
	\includegraphics[width=0.4\textwidth, trim= 2 2 45 30, clip=true] {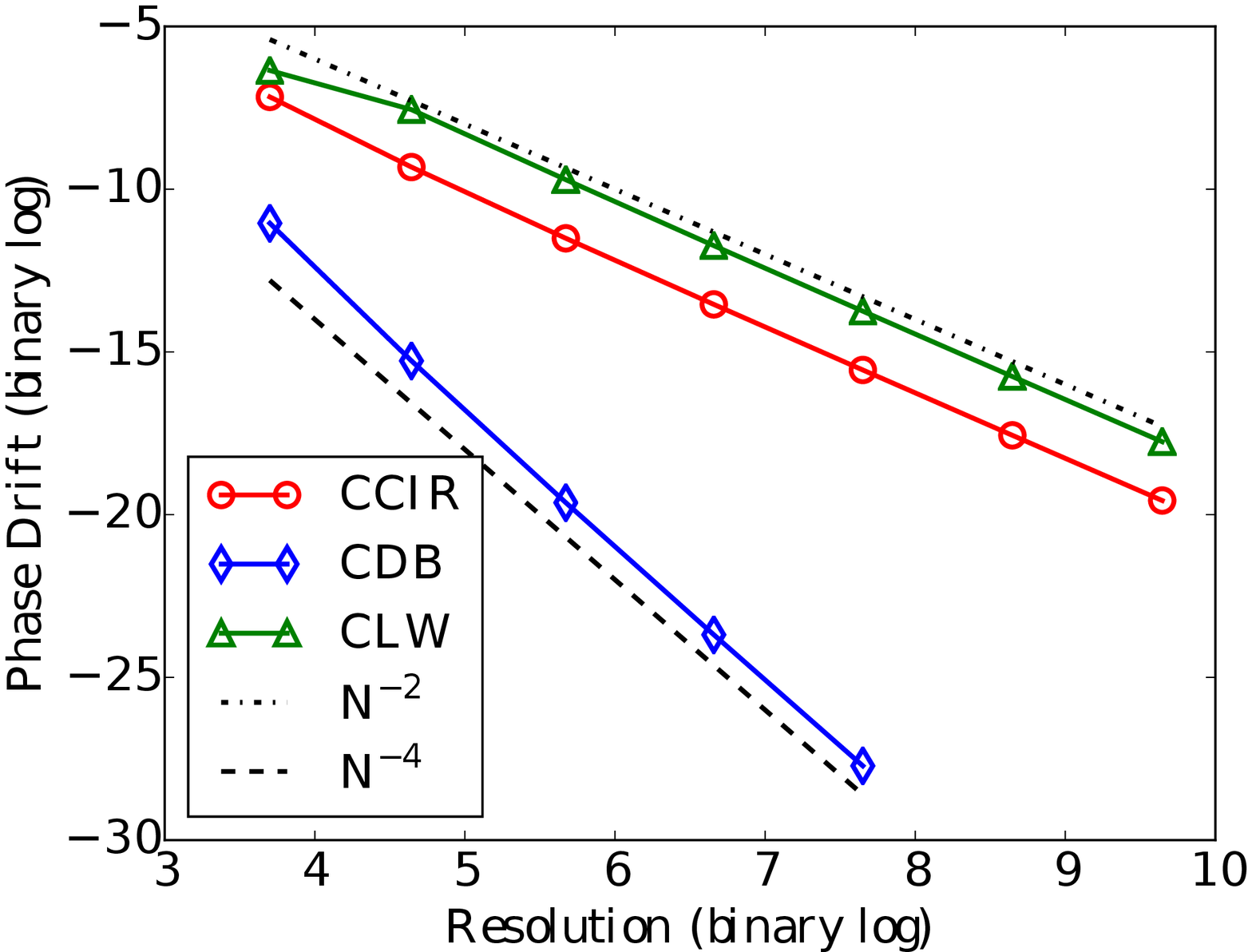}
	}}
	\caption{Decay rate (loss in amplitude) and phase shift per unit of time for the
          test cases presented in fig.~\ref{fig:testHTC} at time $t=5$. The nature of the
          leading order error term (diffusive or dispersive) is clearly highlighted.}
\label{fig:GrwPha1D}
\end{figure}

These schemes are compared in figs.~\ref{fig:mass1D}-\ref{fig:GrwPha1D}. 
First, we consider the evolution of the total mass in a simple test case of
a periodic flow of the form $u=\sin(x)$ with a constant initial distribution
of mass $\Phi(t=0; x)=1$. This is illustrated in fig.~\ref{fig:mass1D}. The 
conservative property of the \CCIR, \CLW{} and \CDB{} schemes is highlighted
by the plot of the total mass which remains constant equal to its initial value. 
Fig.~\ref{fig:testHTC} shows standard tests of advection in a periodic domain 
of a Heaviside, piecewise affine and cosine functions. The diffusive or 
dispersive behavior generated by the order error term are confirmed. The 
order can be quantified with more details by considering the error on the 
amplitude and the phase of the cosine profile 
(\textit{e.g.} \cite{cameron_multi-stage_2016}).
Fig.~\ref{fig:GrwPha1D} illustrates that the order of the original scheme 
is maintained for its conservative counterpart.

\begin{figure}
	\centerline{
	\subfigure[CFL$=0.75$]{
	\includegraphics[width=\tlwidth, trim= 2 2 55 44, clip=true] {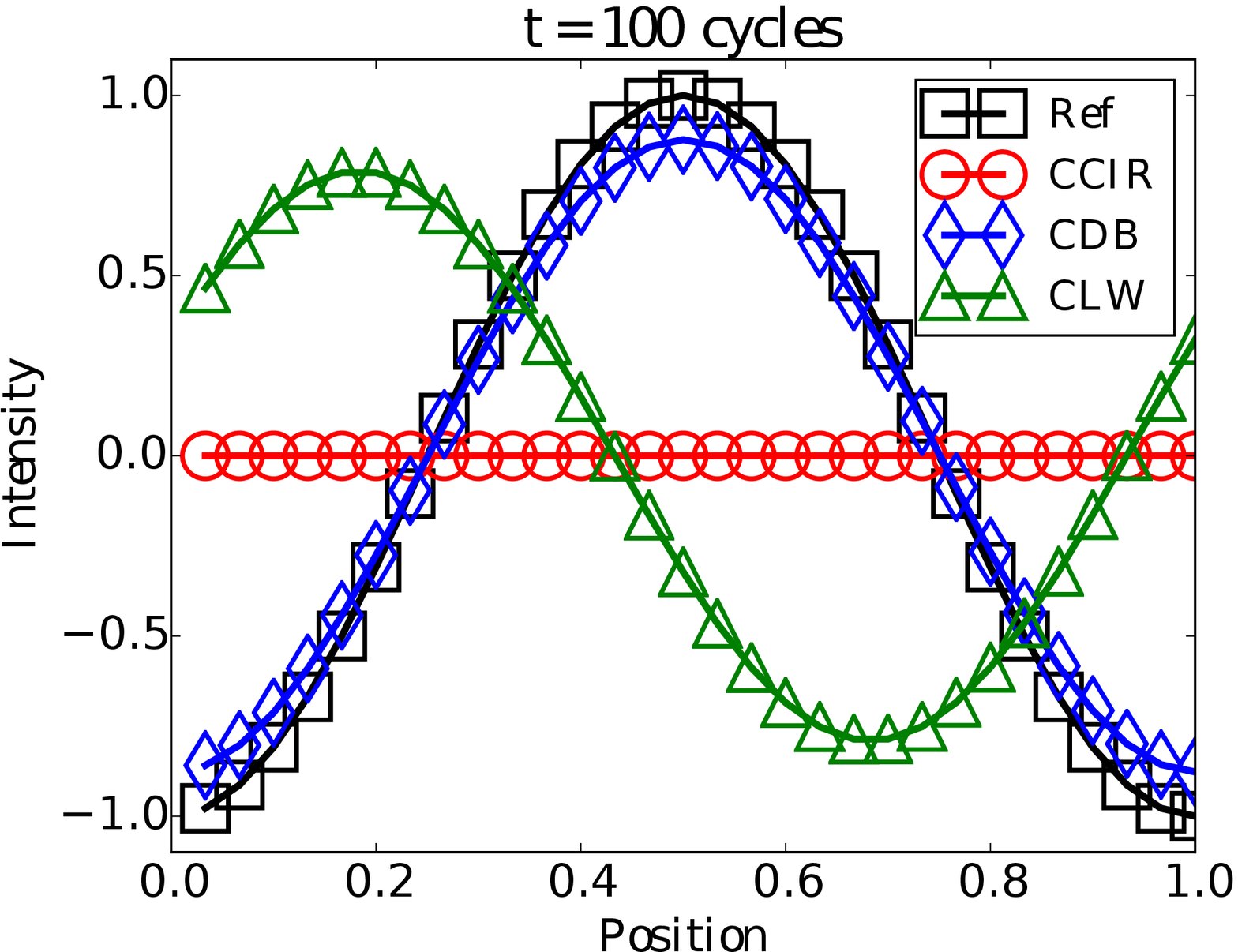} 
	}
	\subfigure[CFL$=2.5$]{
	\includegraphics[width=\twidth, trim= 20 2 54 44, clip=true] {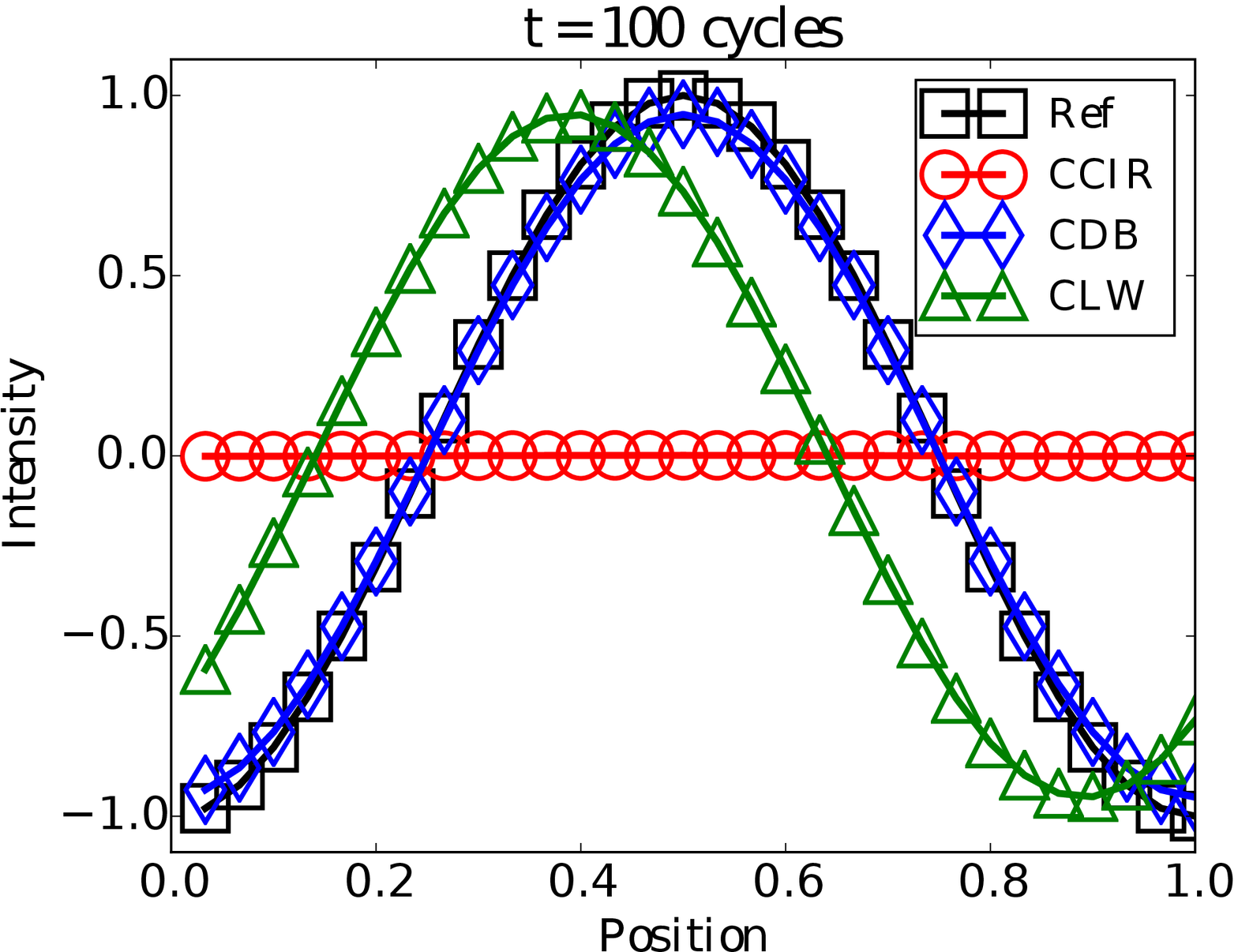} 
	}
	\subfigure[CFL$=7.5$]{
	\includegraphics[width=\twidth, trim= 20 2 54 44, clip=true] {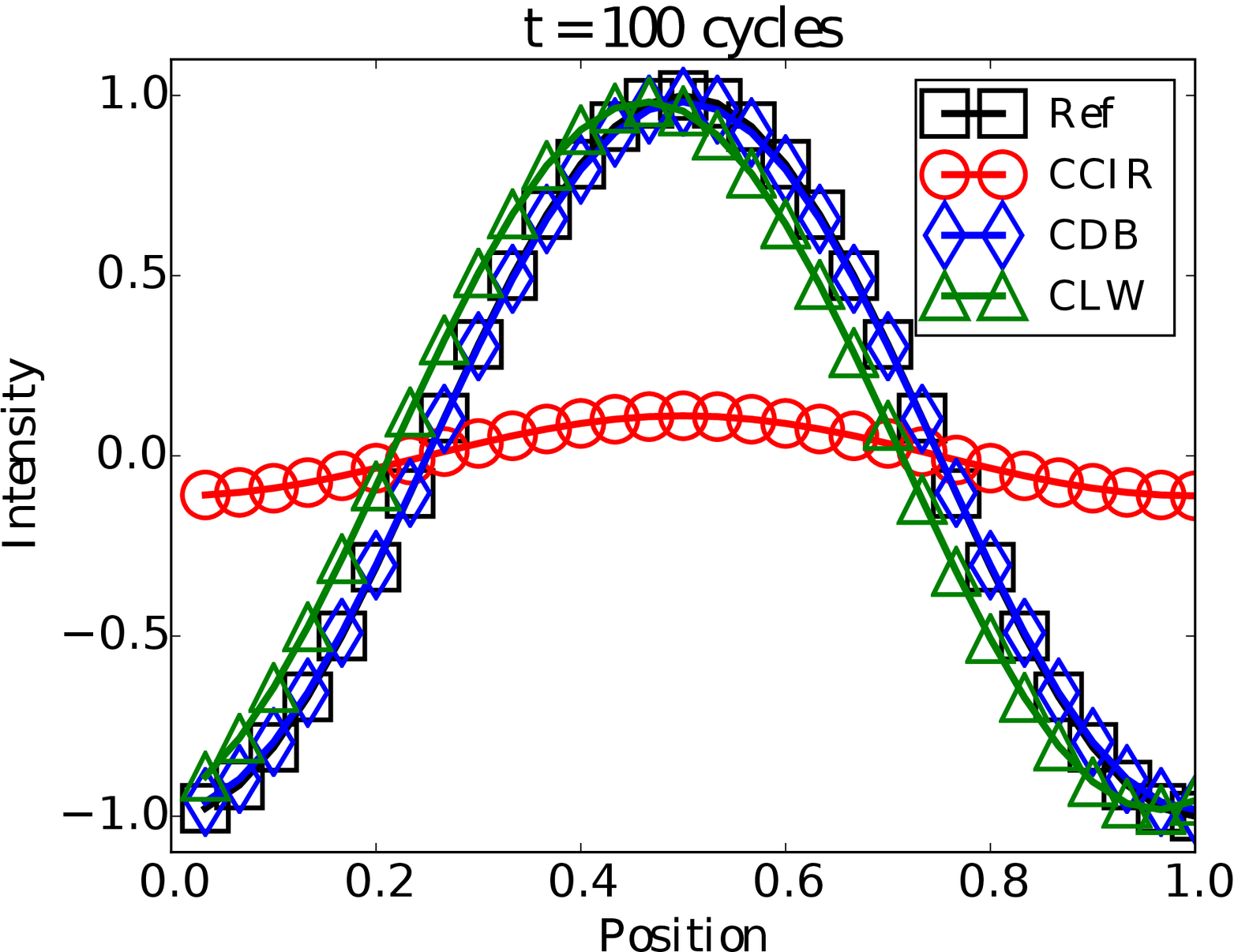}
	}}
	\caption{Advection of a cosine function over $100$ periods of the flow with a
          CFL number of $0.75$ (a), $2.5$ (b), and $7.5$ (c).}
	\label{fig:1dADVcfl}
\end{figure}

In order to generalize theses scheme to CFL numbers greater than unity, 
the interpolation point has to be shifted by a integer number of grid spaces,
using
\begin{align}
	\Ut_i = \left[ ( u_i \Delta t ) / \Delta x \right] \% 1
	\; , \;
	j = i -u_i \Delta t / \Delta x + \Ut_i
	\; , \;
	\Ut^{+}_i &= \max ( \Ut_i ,\, 0 )
	\; , \;
	\Ut^{-}_i = \min ( \Ut_i ,\, 0 ) \,.
	\label{eq:modulo}
\end{align}
For example, the conservative \CIR{} scheme then becomes
\begin{align}
	\Phi^{n+1}_i = \Phi^n_j + \left[ (\Ut^+\Phi^n)_{j-1} 
	- (\vert \Ut\vert \Phi^n)_{j} - (\Ut^-\Phi^n)_{j+1}
	\right] \, .	
\end{align}
Similar expressions follow for the other schemes.
The density profiles of the simulation using CFL number above unity, are presented
in fig.\ref{fig:1dADVcfl} in the case of an initial cosine profile.

\section{Extension in higher dimensions}
\label{sec:geocons}
The standard reconstruction used with the \CIR{} scheme is a bilinear reconstruction. 
It takes the form:
\begin{align}
	\Phi^{n+1}_{i,j} =& [(1-\UUa)(1-\VVa) \Phi^n]_{i,j}
	[\UUa (1-\VVa)]_{i,j} \Phi^n_{\alpha,j} 
	\label{eq:2d:CIR} \\& + \nonumber
	[(1-\UUa)\VVa]_{i,j} \Phi^n_{i,\beta} +
	[\UUa \VVa]_{i,j} \Phi^n_{\alpha,\beta} \,,
\end{align}
where $\VV_i= v_i \Delta t / \Delta x$, $\alpha= i - sign(u_{i,j})$ 
and $\beta= j - sign(v_{i,j})$. 

\begin{figure}
	\centerline{
	\subfigure[Advection sketch]{
		\includegraphics[width=0.4\textwidth, trim= 0 0 0 0, clip=true]
			{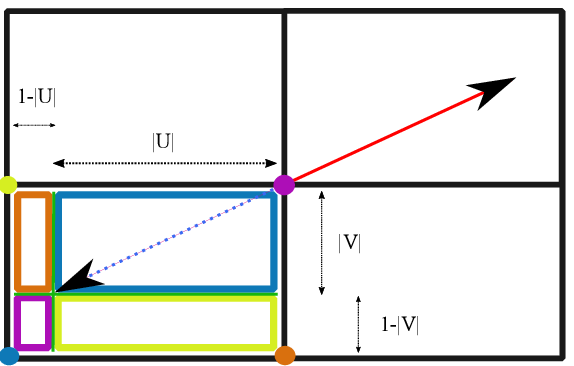}
		\label{fig:adv}
	} \hspace{0.5cm} 
	\subfigure[Continuity sketch]{
		\includegraphics[width=0.4\textwidth, trim= 0 0 0 0, clip=true]
			{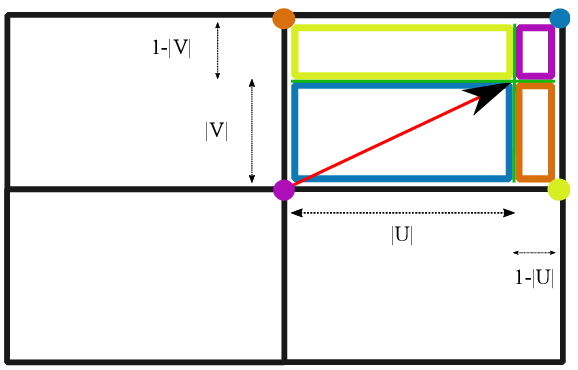}
		\label{fig:csvt}
	} }
	\caption{Illustration of the reconstruction strategy and computational 
	weights for the standard \CIR{} scheme (a) and its conservative \CCIR{} 
	counterpart (b). The red arrow corresponds to the forward advection. 
	The color of a rectangle, indicates its contribution in the evolution 
	of a given point with the same color (see text).} 
	\label{fig:ske}
\end{figure}

The above stencil can be interpreted using the geometric construction
presented in fig.~\ref{fig:adv}. Semi-Lagrangian schemes require to
reconstruct the field at the backward advected points $ {\bf x}_{i,j} -{ \bf u}_{i,j} \Delta t$ . 
Considering a CFL number smaller than unity, the reconstruction point 
necessarily lies in one of the cells surrounding ${\bf x}_{i,j}$. This point 
naturally splits the cell in four parts. The weight of each node in the 
bilinear interpolation eq.~\eqref{eq:2d:CIR} corresponds to the ratio of the 
surface of the rectangle opposite to this node normalised by to the total 
surface of the computational cell. This graphical interpretation of 
eq.~\eqref{eq:2d:CIR} is illustrated on fig.~\ref{fig:ske}.a the backward 
displacement \( -{ \bf u}_{i,j} \Delta t \) being indicated with a dashed line.

Let us now turn to the conservative scheme, the two-dimensional version 
of the \CCIR{} scheme can be expressed as:
\begin{align}
	\nonumber
	\Phi^{n+1}_{i,j} = & 
	[\UUap \VVap \Phi^n]_{ i -1, j -1} + 
	[\UUap (1-\VVa) \Phi^n]_{ i -1 , j} +	
	[\UUap \VVam \Phi^n]_{ i -1, j + 1} 
	\\& +
	[(1-\UUa) \VVap \Phi^n]_{i, j -1 } +
	[(1-\UUa)(1-\VVa) \Phi^n]_{i,j} +
	[(1-\UUa) \VVam \Phi^n]_{i,j + 1 } 
	\label{eq:2d:MSL}
	\\& \nonumber +
	[\UUam \VVap \Phi^n]_{i + 1, j -1} +
	[\UUam (1-\VVa) \Phi^n]_{i + 1 , j} +
	[\UUam \VVam \Phi^n]_{i + 1, j + 1}
	\, .
\end{align}
It is enlighting to interpret this formula geometrically. The weights now 
correspond to the forward displacement \( { \bf u}_{i,j} \Delta t \), indicated
with a solid line on fig.~\ref{fig:ske}.b. Again the weight of each
term is given by the relative surface of the rectangle opposite to the
advected vertex, normalized by to the total surface of the
computational cell. The key distinction is however that the computed
weight corresponds to the contribution of $\Phi_{i,j}$ to the time
evolution of its neighbors. This contrasts with the \CIR{} scheme, for
which the computed weights correspond to the contribution of each 
neighbor to the evolution of $\Phi_{i,j}$.

In fig.~\ref{fig:ske}, mass conservation appears as a direct consequence 
of the fact that the sum of each sub-rectangle amounts to the total 
cell as highlighted by expression \eqref{eq:2d:MSL}. Let us stress that this 
approach results in a conservative non-split semi-Lagrangian formulation.

A few observations can be made on this stencil. First, this rather simple 
geometric interpretation can be generalized to higher dimensions. 
Second, the two-dimensional \CCIR{} stencil of eq.~\eqref{eq:2d:MSL} is 
identical to the split formula corresponding to the composition of two one-dimensional
\CCIR{} stencils, $\CCIR_{xy} = \CCIR_x \circ \CCIR_y = \CCIR_y \circ \CCIR_x $.
Such is not the case for the \CIR{} stencil. This commuting property can be used to 
generalize the higher-order conservative schemes from section~\ref{sec:consSL} 
to higher dimensions of space. 

\begin{figure}
	\centerline{
	\subfigure[Initial density profile]{
	\includegraphics[width=0.4\textwidth, trim= 45 25 75 35, clip=true] {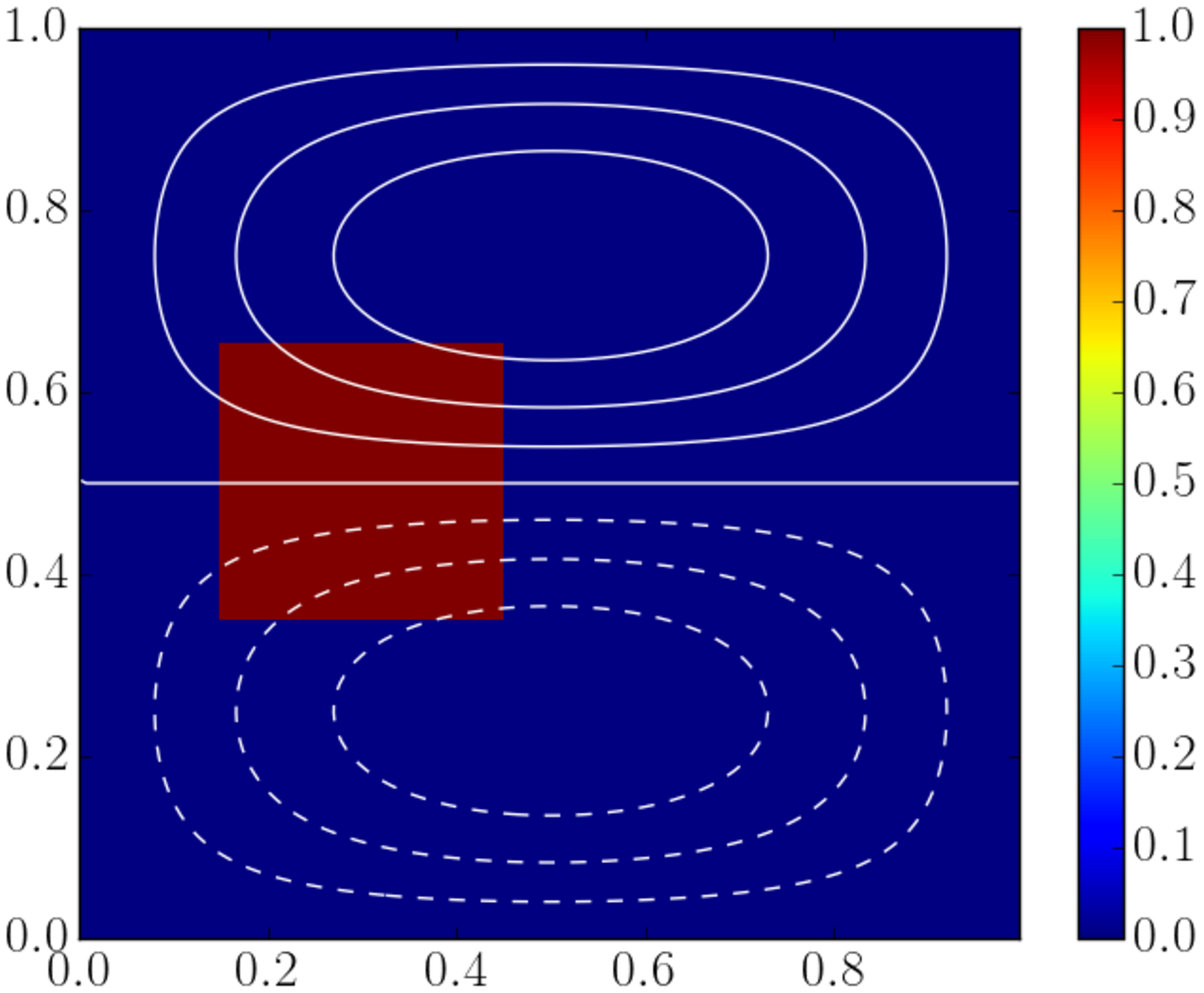}
	} \hspace{0.5cm} 
	\subfigure[Mass evolution with time.]{
	\includegraphics[width=0.4\textwidth, trim= 15 0 45 30, clip=true] {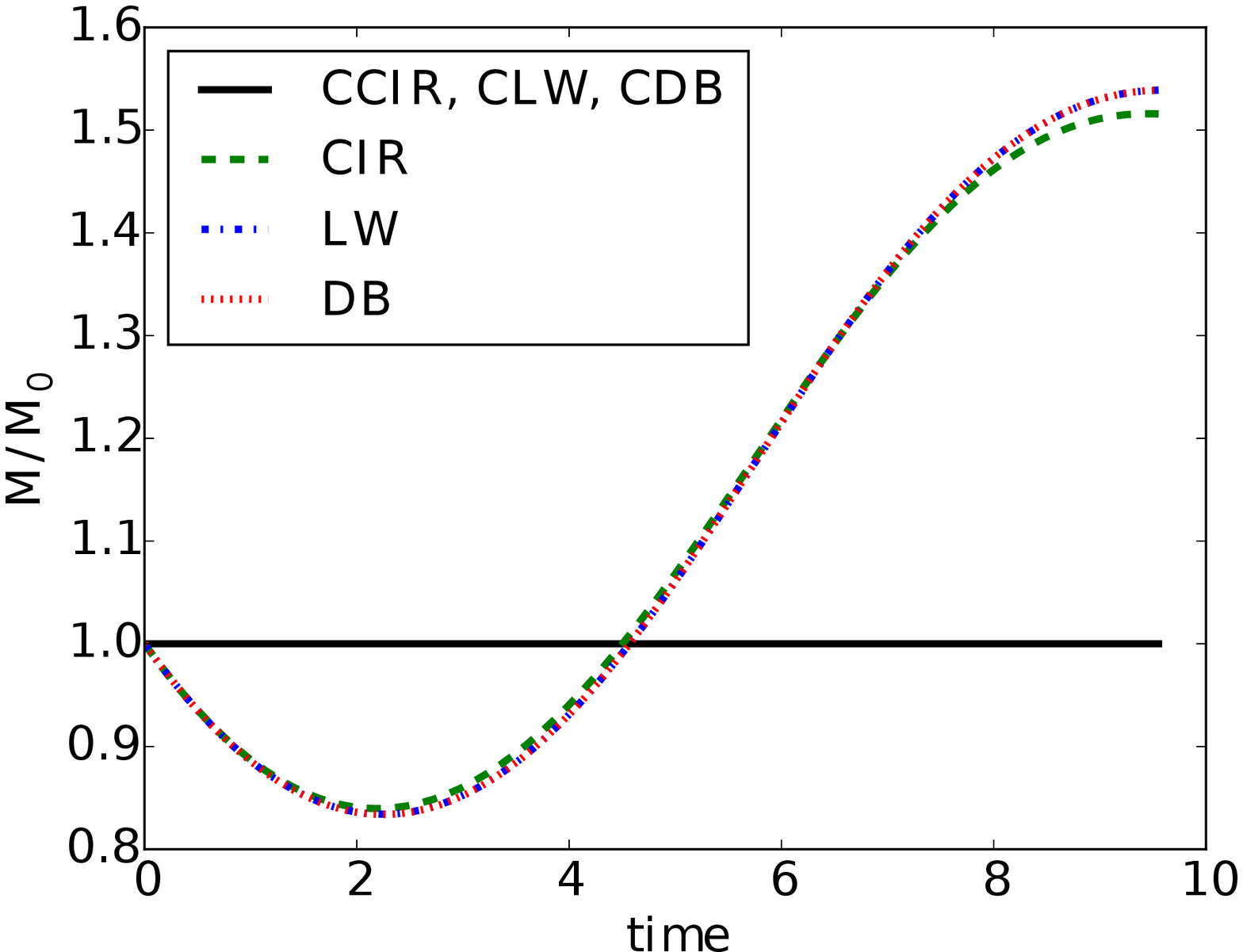}	
	}}
\caption{Two dimensional transport of a density distribution initially uniform 
within a square (a). The total mass evolution with time up to $t=10$ with a 
resolution of $128^3$ for conservative and non-conservative semi-Lagrangian 
schemes of first, second and third order (b) reflects the conservative nature of the schemes.}
\label{fig:patch2D}
\end{figure}

To illustrate the conservative property of the \CCIR{} scheme
in two dimensions of space, it was tested using an incompressible 
velocity profile of the form $ u(t;x,y) = - \sin( \pi x ) \cos ( 2\pi y )$, 
$v(t;x,y) = \cos( \pi x ) \sin( 2\pi y )$. The initial passive scalar field takes 
the form of a uniform patch $\Phi (t=0; x,y) = 1$ if 
$\vert x-0.5 \vert \leq 0.15$ and $\vert y-0.3 \vert \leq 0.15\, ,$ 
and $0$ elsewhere (see fig.~\ref{fig:patch2D}.a).

Since the flow is incompressible, the advection and continuity equation 
are equivalent. We thus compare the three schemes discussed in
section~\ref{sec:consSL} and their conservative counterpart.
In fig.~\ref{fig:patch2D}b, the evolution of relative total mass 
of the \CIR{}, \LW{} and \DB{} schemes is represented. As expected, 
the conservative schemes have a relative mass
equal to unity, up to machine precision for the same set of parameters.

\begin{figure}
	\centerline{
	\ (a)\includegraphics[height=\theight, trim= 45 25 130 35, clip=true] {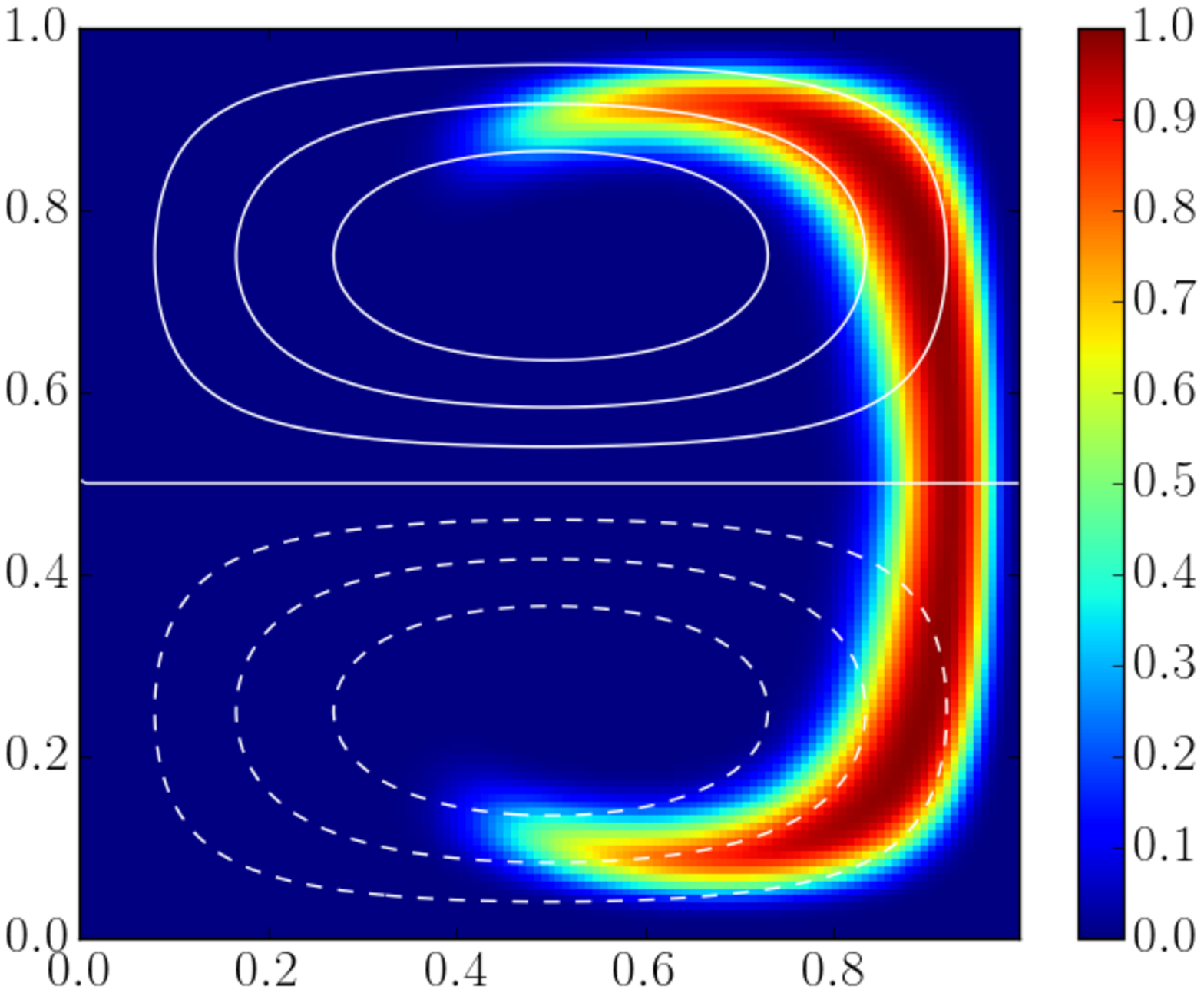}
	\ (b)\includegraphics[height=\theight, trim= 45 25 130 35, clip=true]{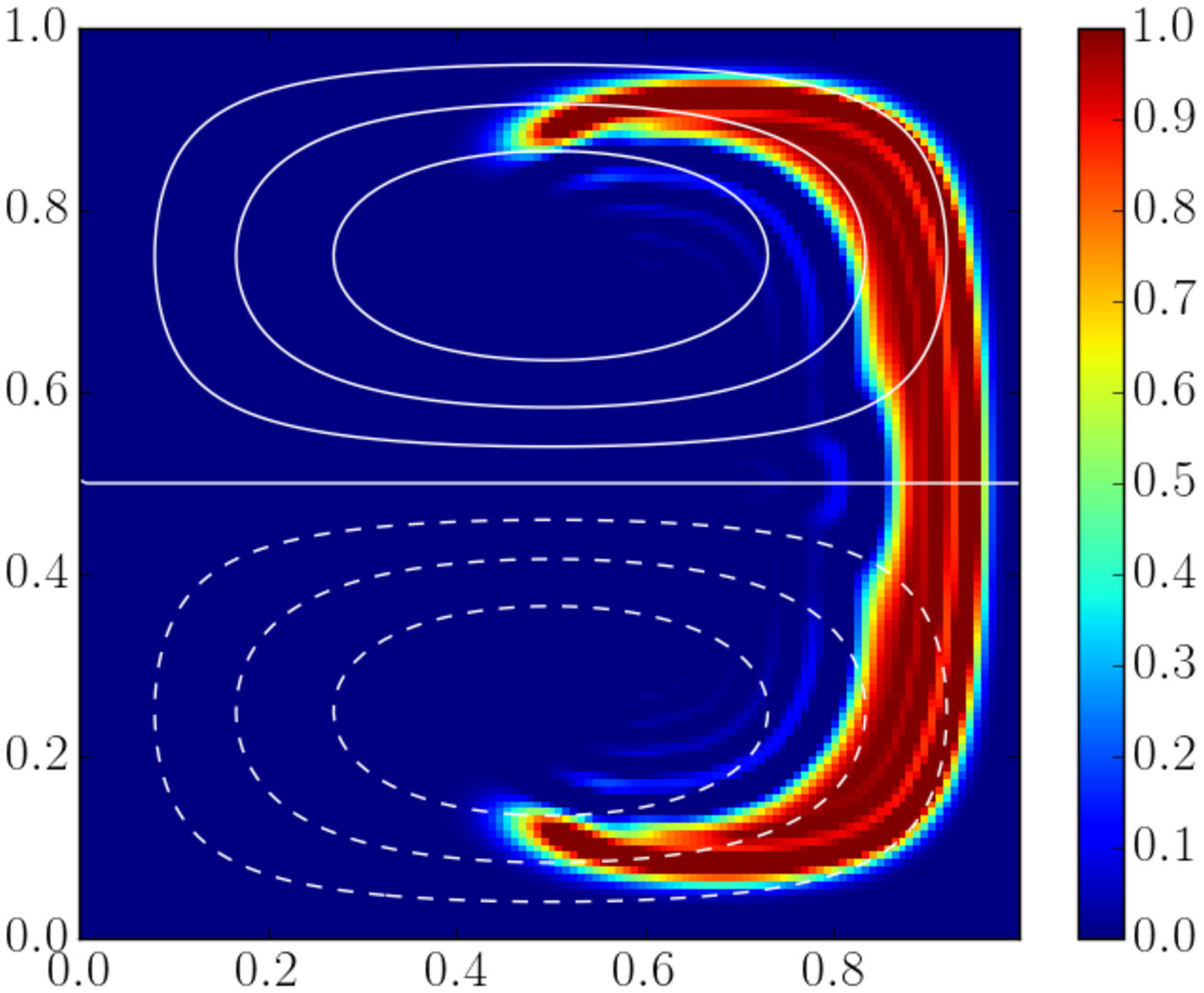}
	\ (c)\includegraphics[height=\theight, trim= 45 25 75 35, clip=true] {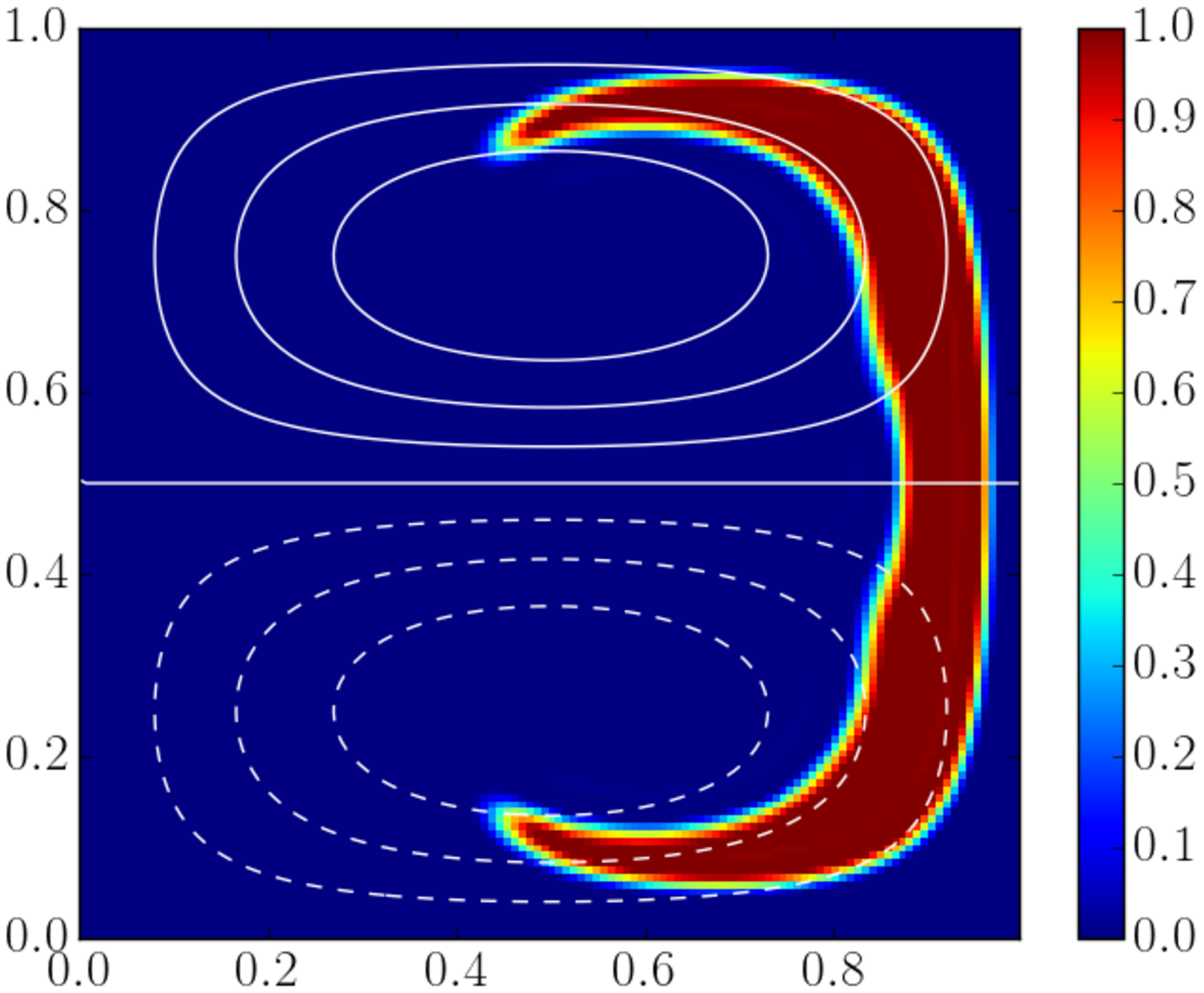}
	}
	\centerline{
	\ (d)\includegraphics[height=\theight, trim= 45 25 130 35, clip=true] {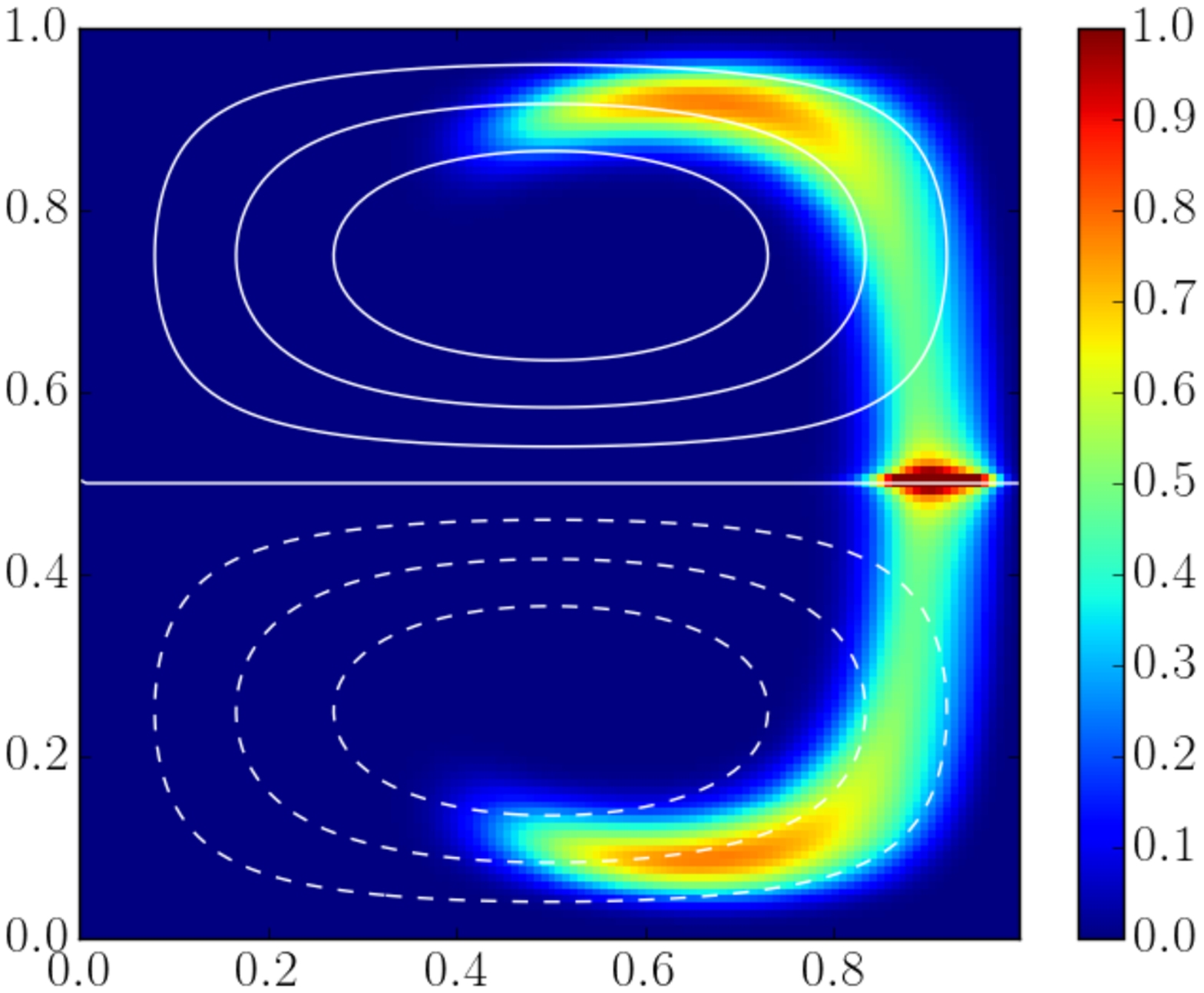}
	\ (e)\includegraphics[height=\theight, trim= 45 25 130 35, clip=true] {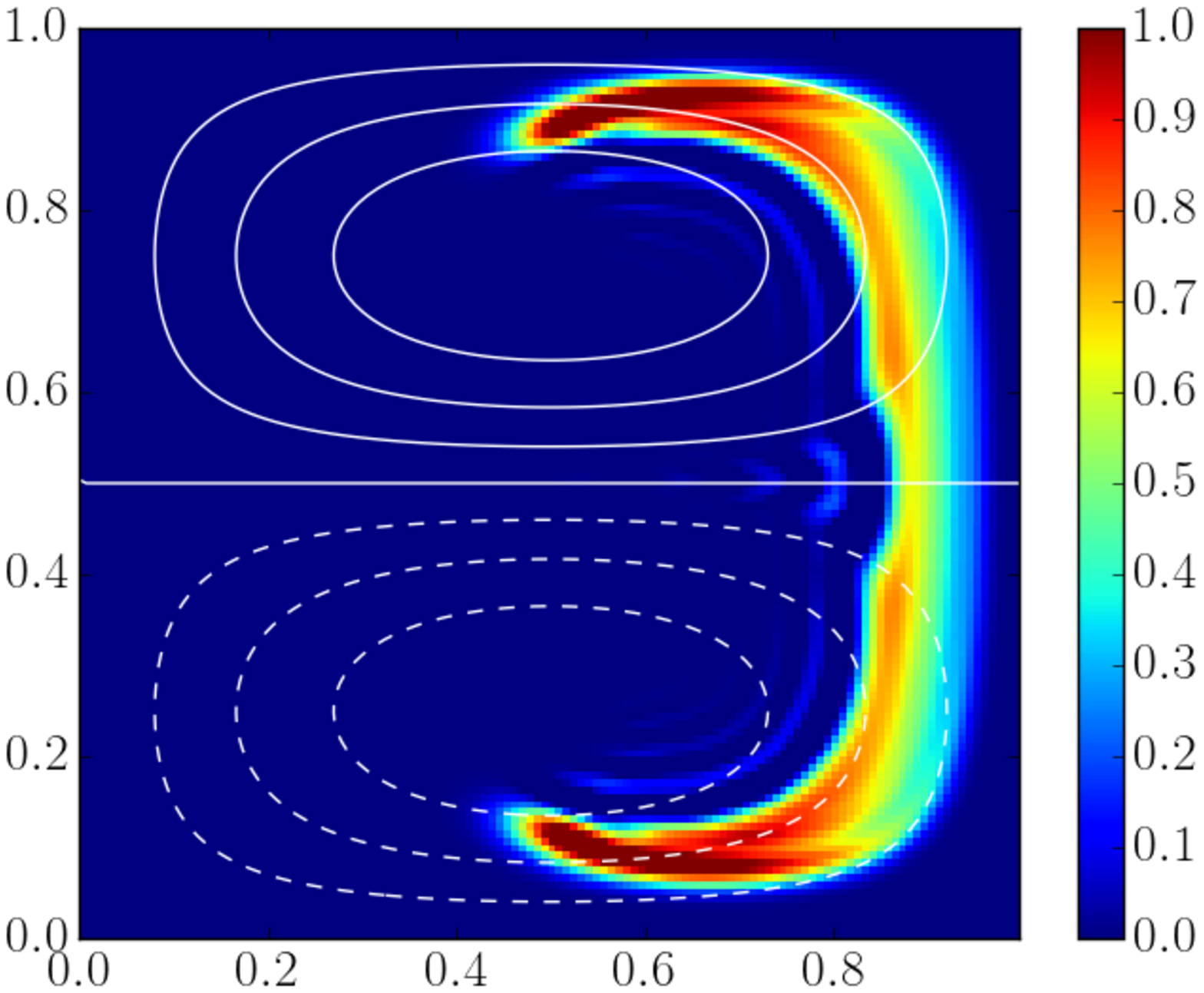}
	\ (f)\includegraphics[height=\theight, trim= 45 25 75 35, clip=true] {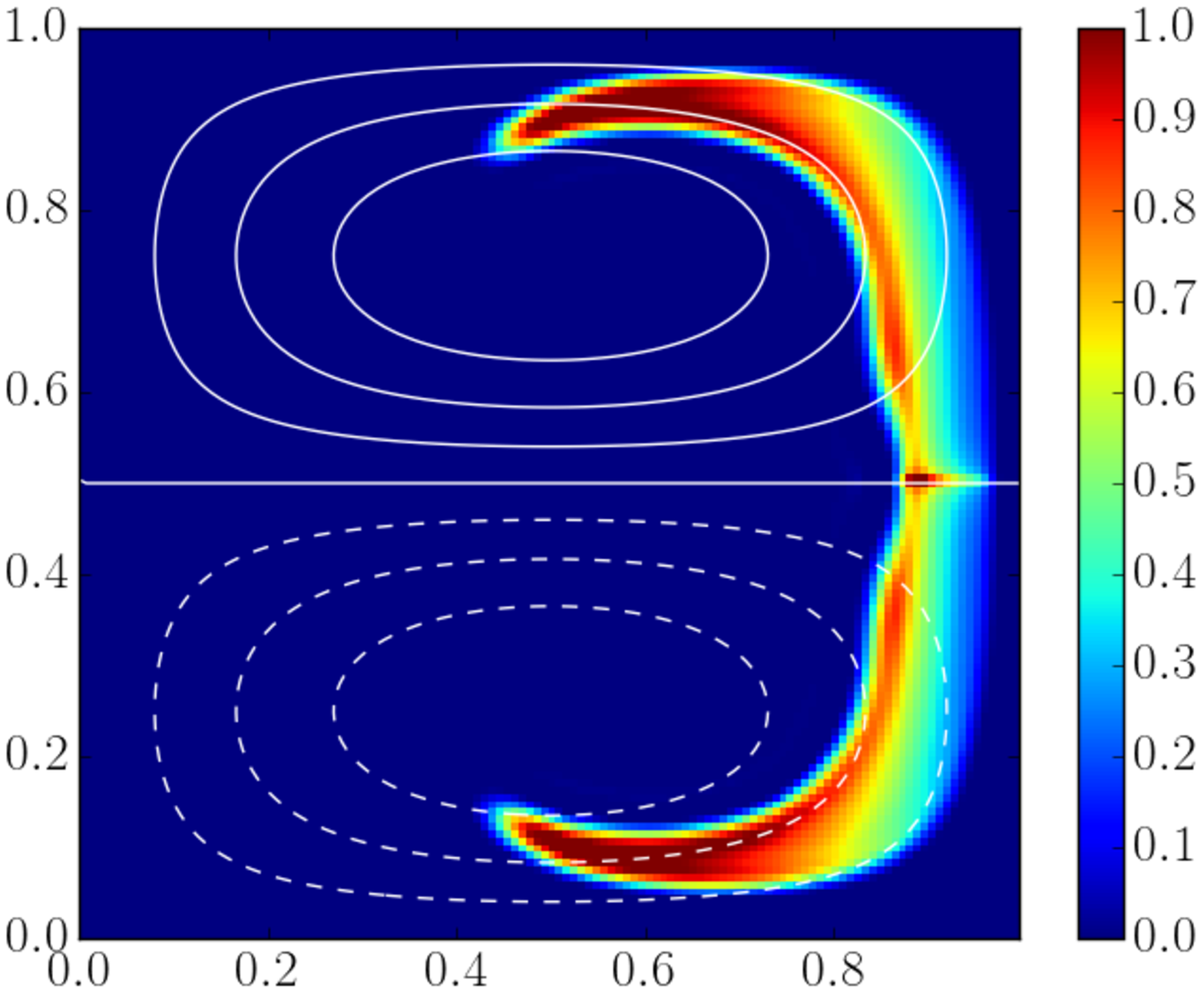}
	}
	\centerline{
	\ (g)\includegraphics[height=\theight, trim= 45 25 130 35, clip=true] {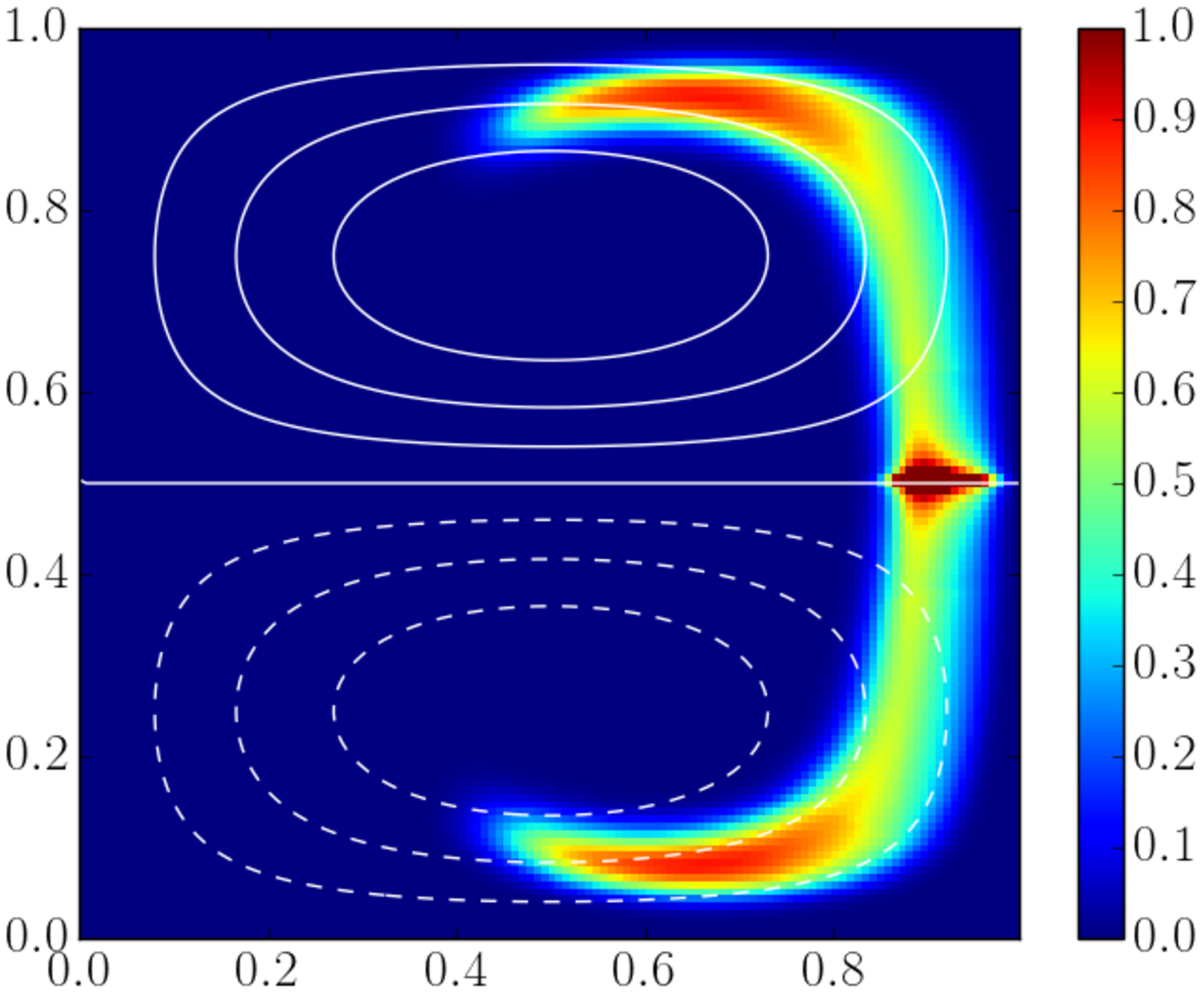}
	\ (h)\includegraphics[height=\theight, trim= 45 25 130 35, clip=true] {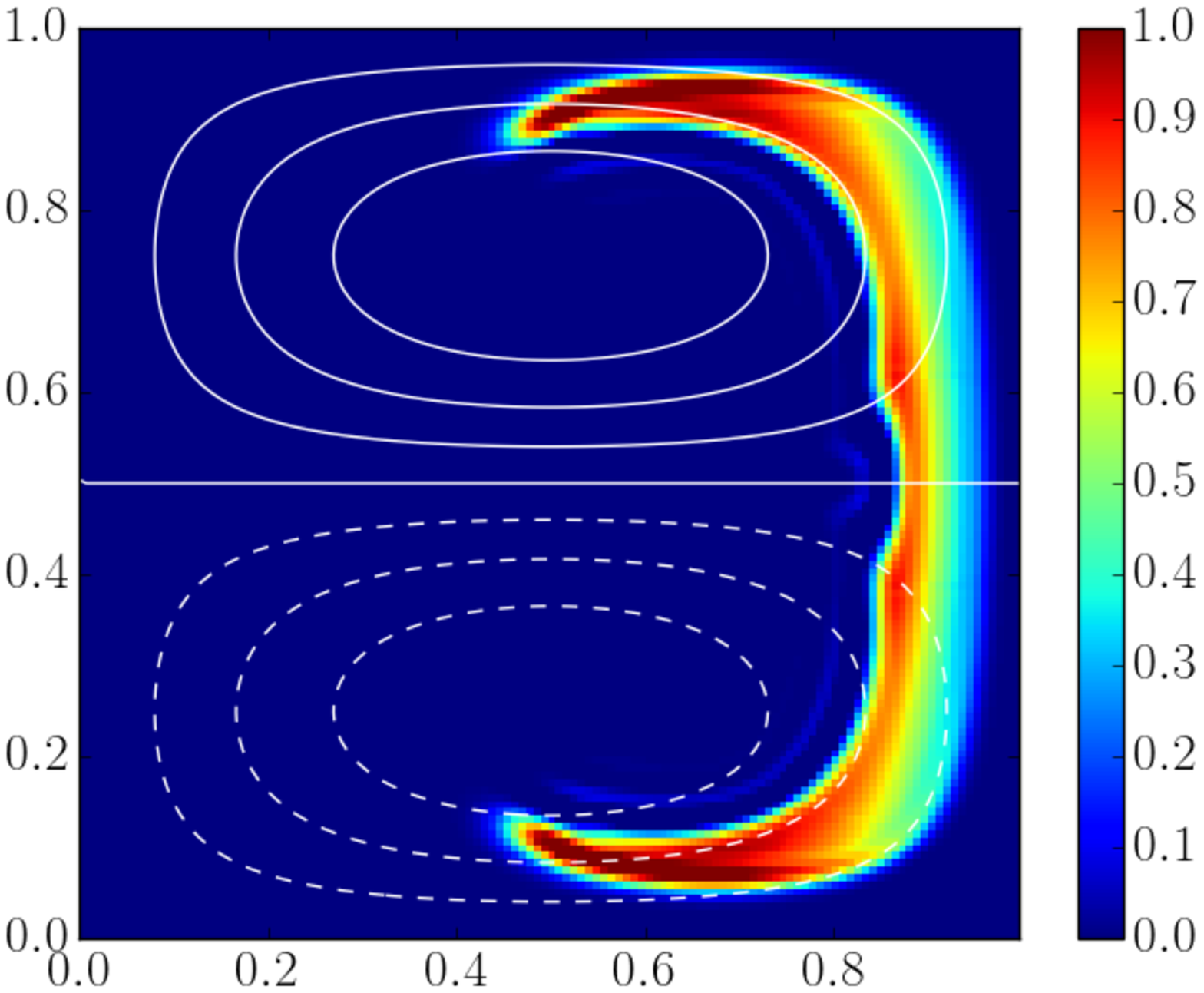}
	\ (i)\includegraphics[height=\theight, trim= 45 25 75 35, clip=true] {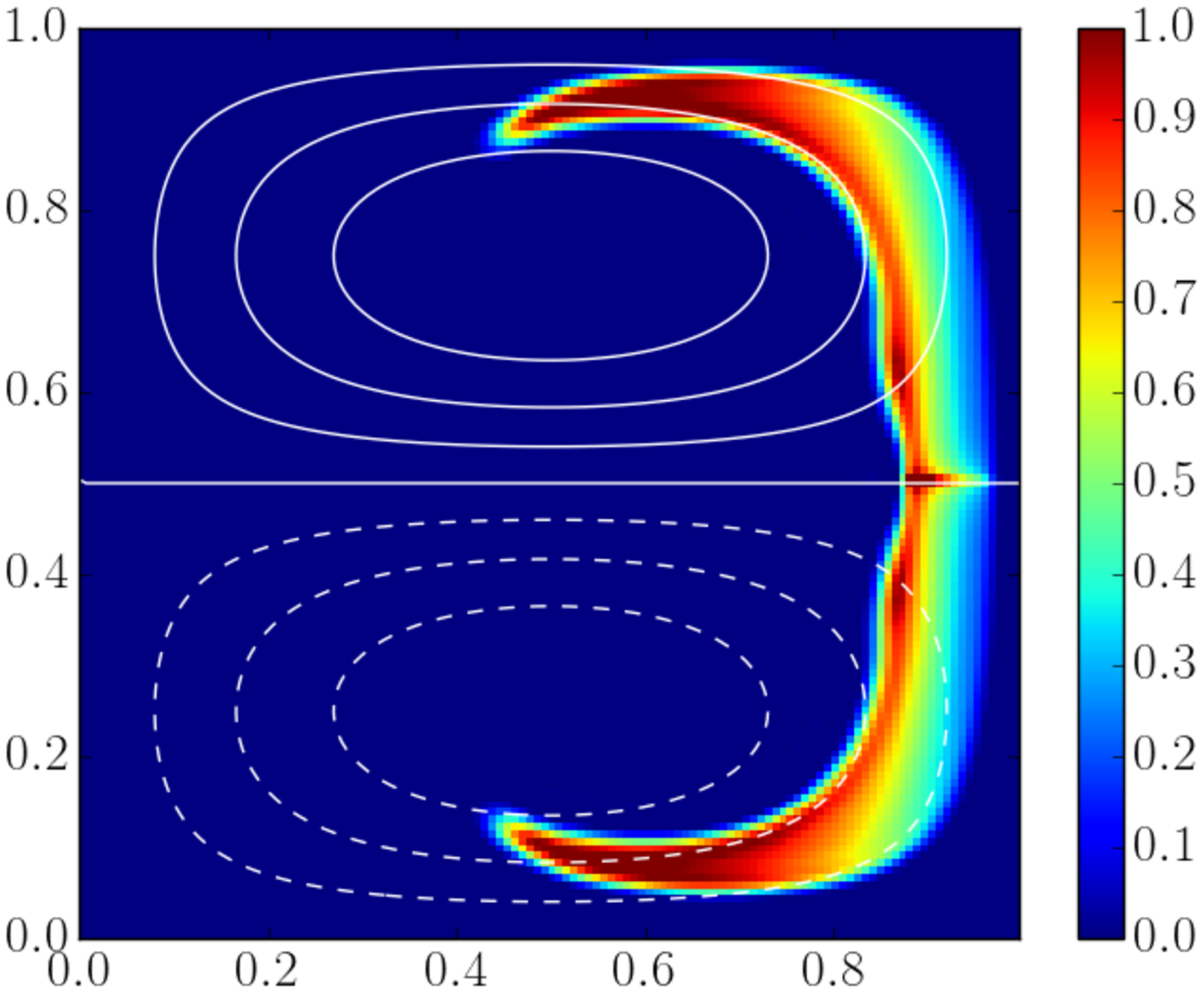}
	}
    \caption{A comparison of the non-conservative scheme at CFL$ = 0.8$ (a,b,c)
    with the conservative scheme at CFL$=0.8$ (d,e,f)
    and at CFL$=1.6$ (g,h,i). The simulation were carried out
    at a resolution of $128^2$ for an integration time of $t=10$.}
    \label{fig:patch2Dt}
\end{figure}

In fig.~\ref{fig:patch2Dt}, color-plots of the density profile are given 
for all schemes at $t=10$. The evolution of mass in the plan of symmetry 
is different for all schemes. The accumulation of mass near the 
stagnation point is clearly visible with the conservative schemes of odd 
orders, see figs.~\ref{fig:patch2Dt}(d) and \ref{fig:patch2Dt}(f). Dispersive 
effects in fig.~\ref{fig:patch2Dt}(e), which do not vanish in the symmetry 
plane, are still too strong to allow for this feature to emerge. 

The explicit scheme introduced in eq.~\eqref{eq:2d:MSL} corresponds to a
first order time integration. We should stress however that the modified reconstruction 
strategy introduced to enforced conservativity only concerns the spatial operator.
The conservative property is thus retrained for higher order or multi-level time-stepping
algorithms as shown in eqs.\eqref{eq:CNeq} and \eqref{eq:CCNeq}.

\begin{figure}
	\centerline{
	\ (a)\includegraphics[height=\theight, trim= 45 25 130 35, clip=true] {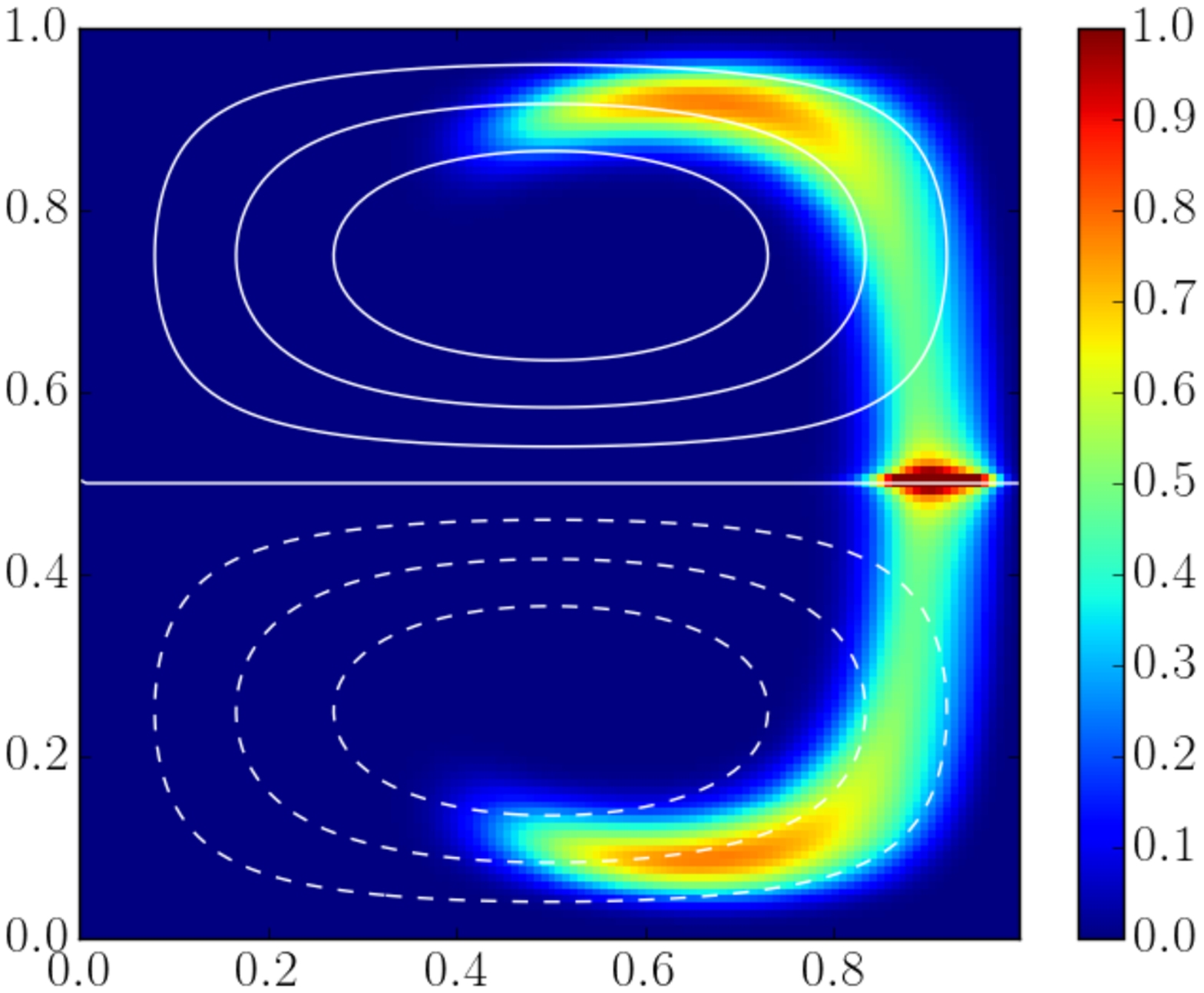}
	\ (b)\includegraphics[height=\theight, trim= 45 25 130 35, clip=true] {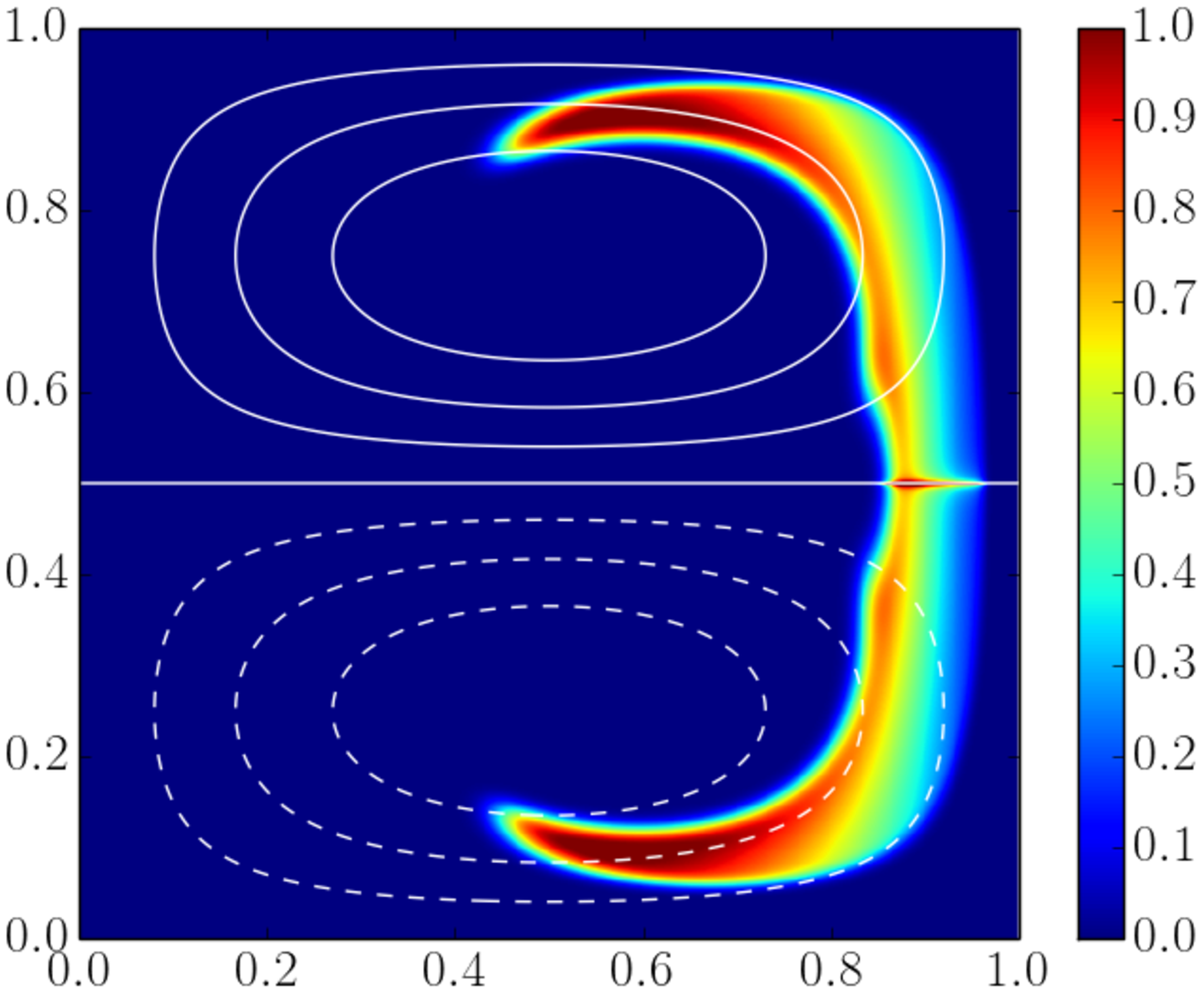}
	\ (c)\includegraphics[height=\theight, trim= 45 25 75 35, clip=true] {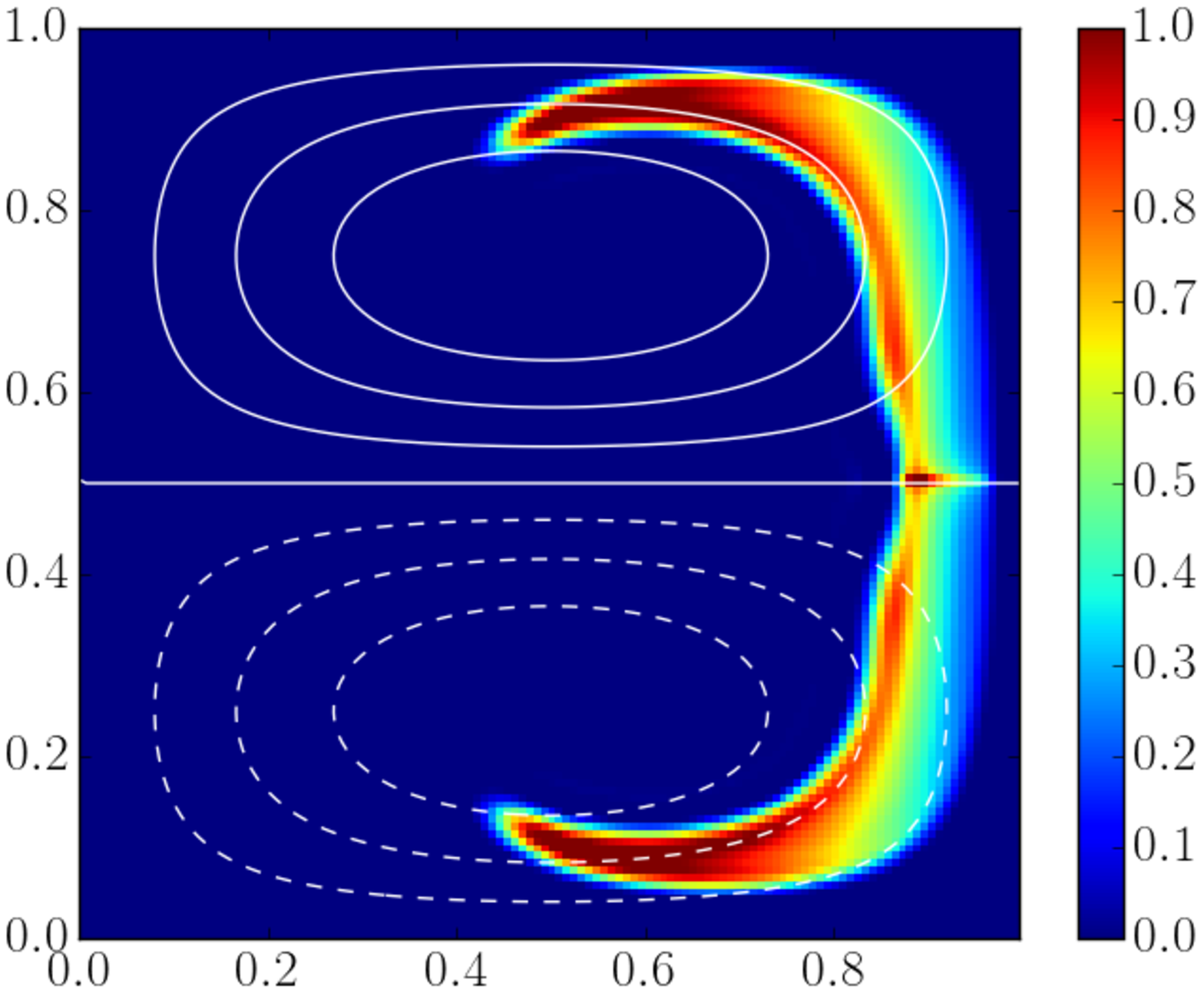}
	}
	\caption{Comparison of the first order conservative \CCIR{} scheme with the third order 
	conservative \CDB{} scheme at CFL$=0.8$. Plots (a) and (b) compare simulations of 
	resolution $128^2$ and $1024^2$ respectively for the \CCIR{} scheme; plot (c) 
	presents the same setup solved with the \CDB{} scheme at a resolution 
	of $128^2$.}
\label{fig:2dCONV}
\end{figure}

In fig.~\ref{fig:2dCONV}, convergence effects can clearly be identified by 
comparing results obtained with the \CCIR{} scheme (conservative, first order) 
with a fine grid ($1024^2$), to the ones obtained with a coarser grid ($128^2$) 
or with the \CDB{} scheme (conservative, third order). At low resolution, owing 
to the effects of the numerical diffusion, the density on fig.~\ref{fig:2dCONV}(a) 
appears to be spread across three independent lobes. Increasing the resolution, 
or using a higher-order scheme, reveals the fine filaments of mass connecting 
these lobes.

\begin{figure}
	\centerline{
	\ (a)\includegraphics[height=\theight, trim= 45 25 130 35, clip=true] {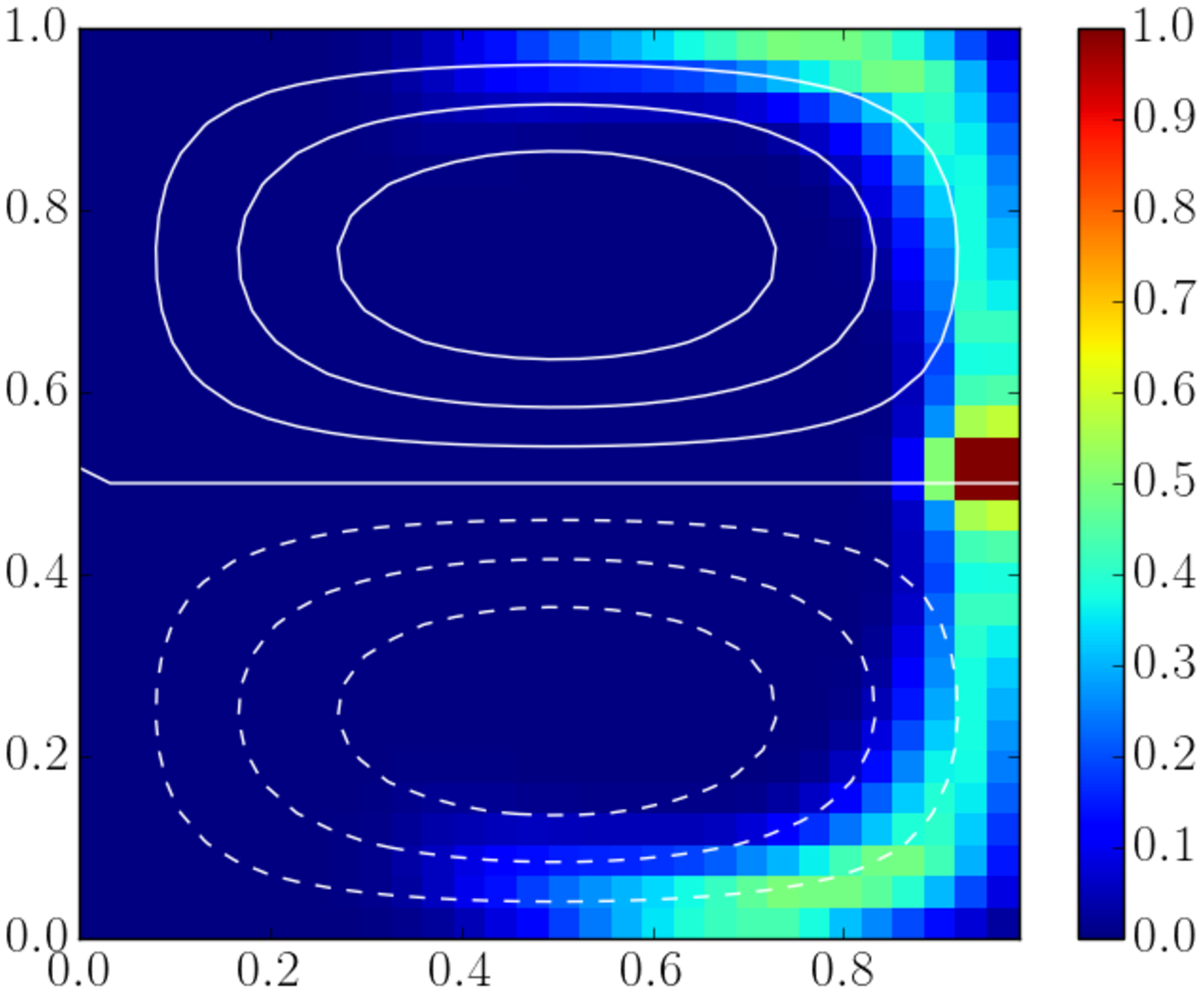}
	\ (b)\includegraphics[height=\theight, trim= 45 25 130 35, clip=true] {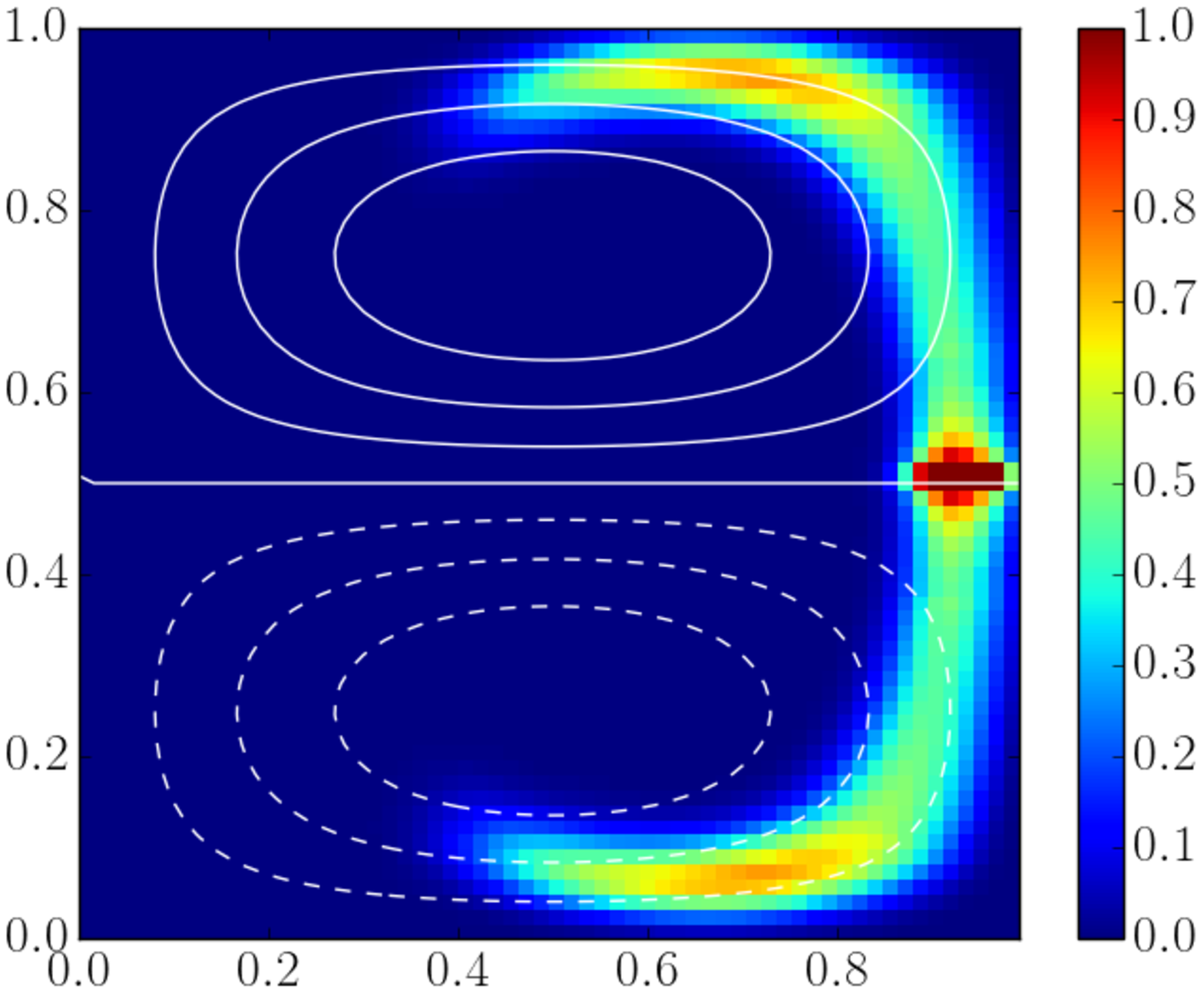}
	\ (c)\includegraphics[height=\theight, trim= 45 25 75 35, clip=true] {im/16e-1_CCIR_0128.eps}
	}\centerline{
	\ (d)\includegraphics[height=\theight, trim= 45 25 130 35, clip=true] {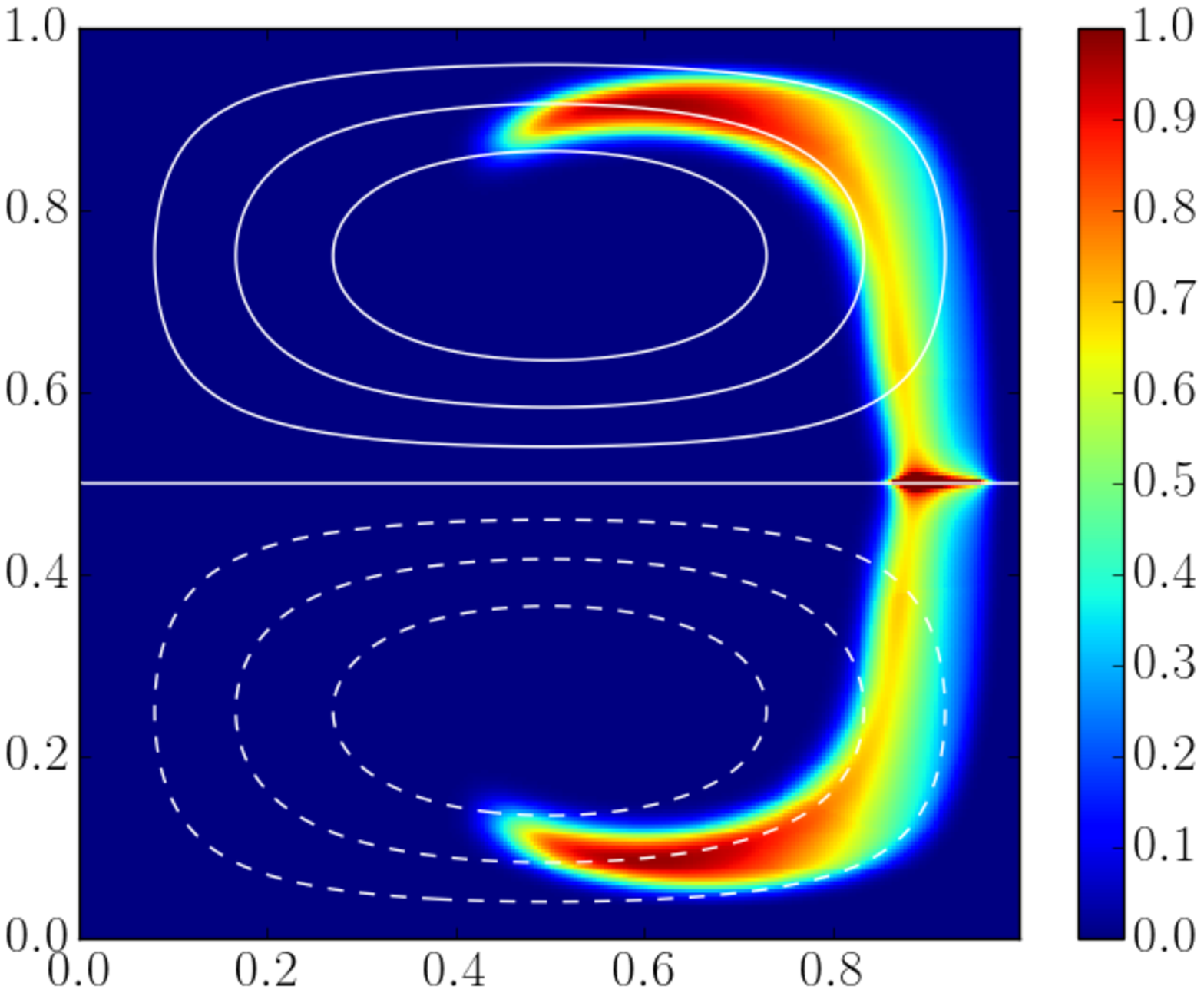}
	\ (e)\includegraphics[height=\theight, trim= 45 25 130 35, clip=true] {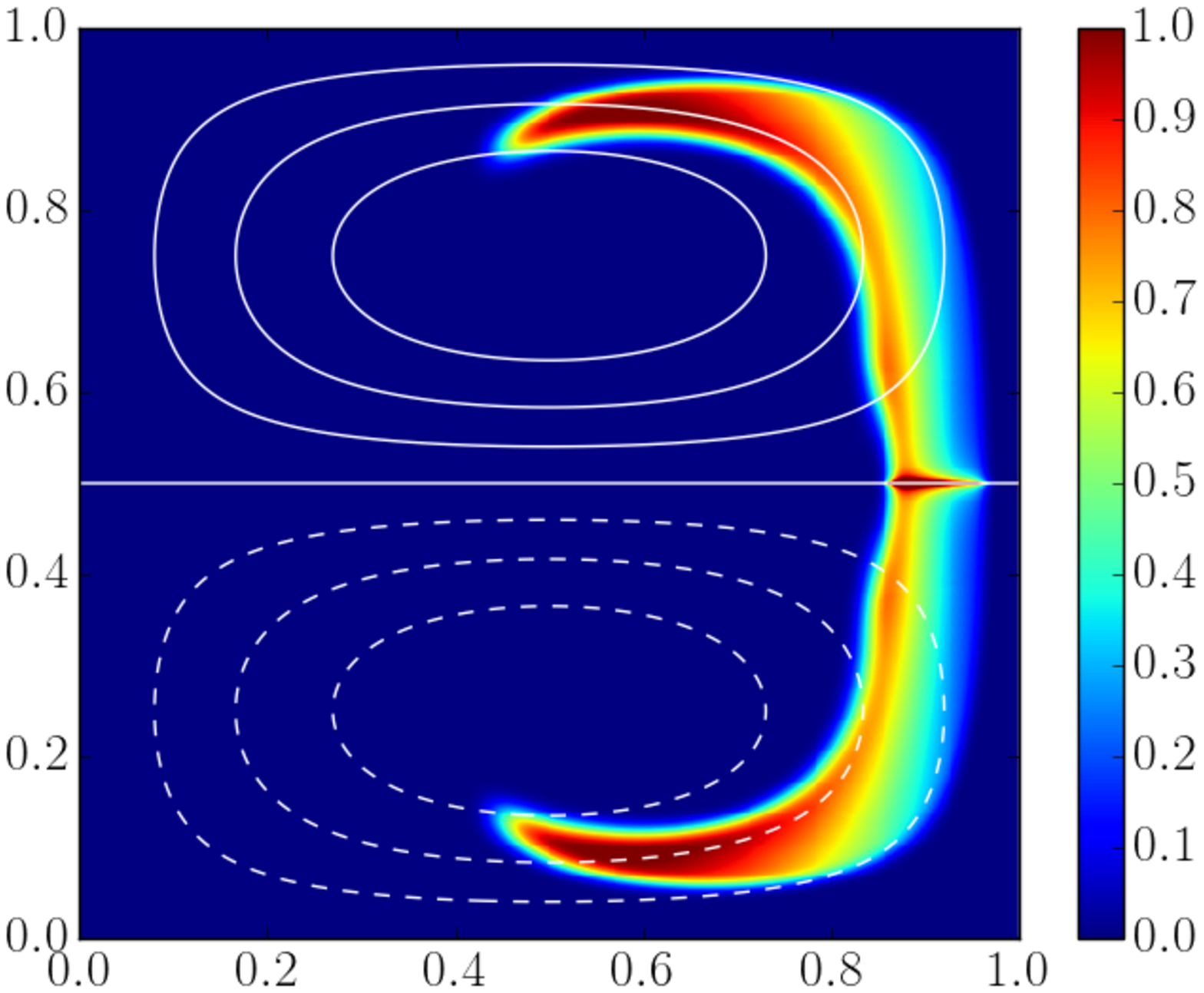}
	\ (f)\includegraphics[height=\theight, trim= 45 25 75 35, clip=true] {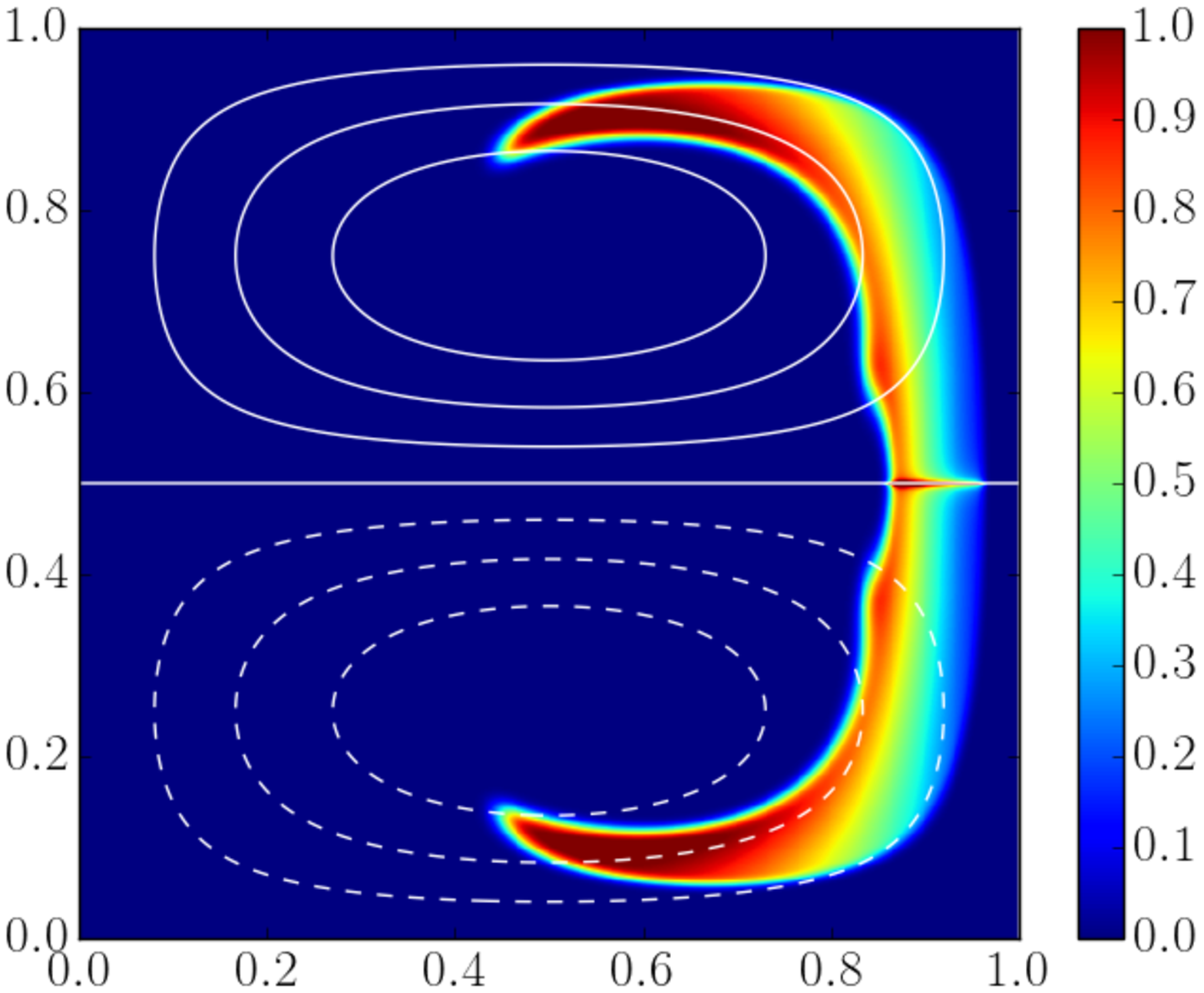}
	}
	\caption{Convergence study for the first order conservative \CCIR{} 
	at CFL$=1.6$ with resolutions:
	$(a):32^2\, , \ (b):64^2 \, , \ (c):128^2 \, , \ (d):256^2 \, , \ (e):512^2 \, , \ (f):1024^2 \, .$ 
	}
\label{fig:2dCFL16e-1}
\end{figure}

Varying the resolution, the convergence of the density profile is tested for the conservative 
diffusive monotone \CCIR{} scheme at CFL$=1.6$ in fig.~\ref{fig:2dCFL16e-1}.
As the resolution increases at constant CFL, the numeric error decreases and 
the density profile becomes closer to the analytic solution. At high resolution, 
the grid is finer, the simulation is therefore more precise and catches 
the details of the structure near the symmetry axis. 


\begin{figure}
	\centerline{
	\ (a)\includegraphics[height=\theight, trim= 45 25 130 35, clip=true] {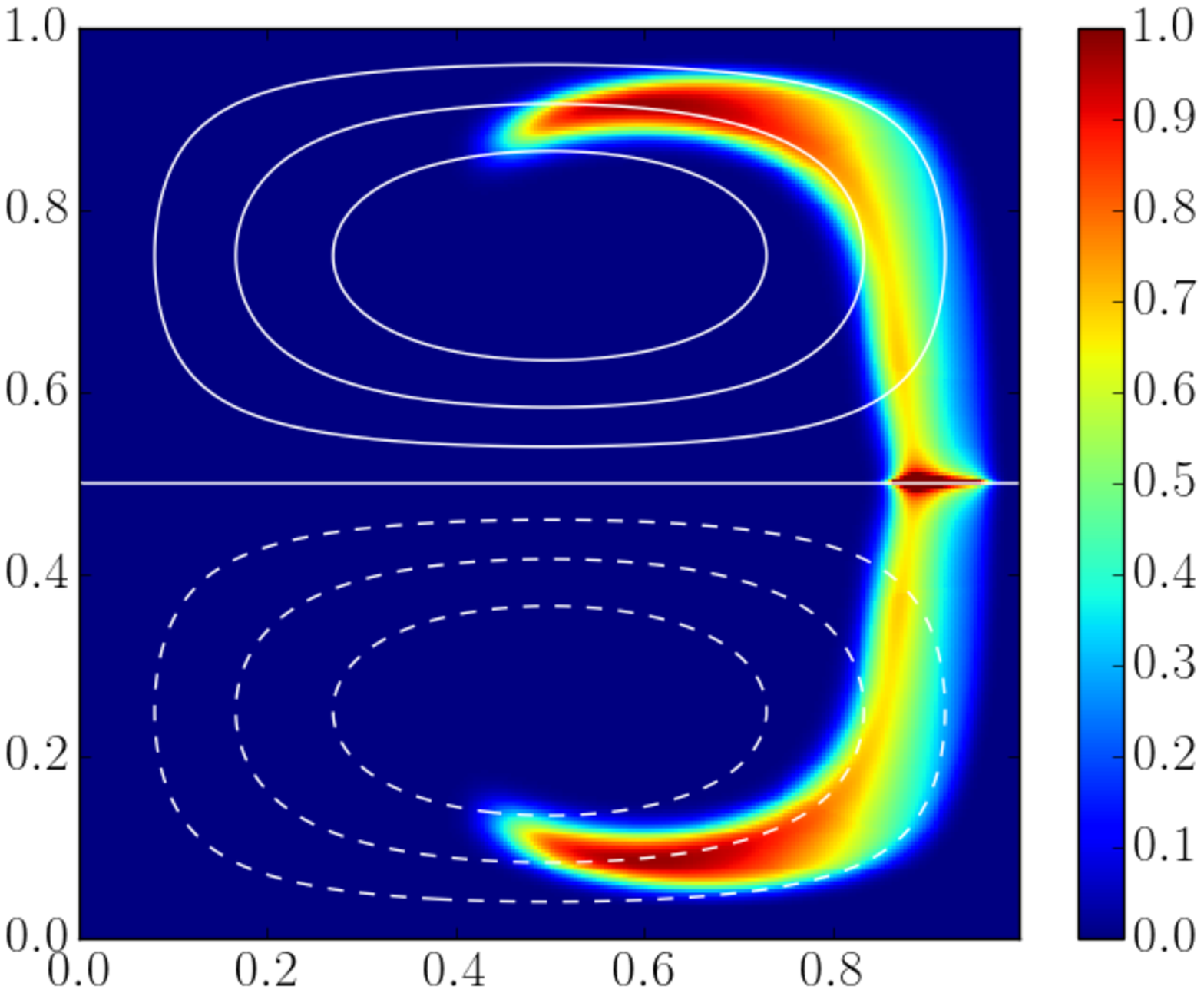}
	\ (b)\includegraphics[height=\theight, trim= 45 25 130 35, clip=true] {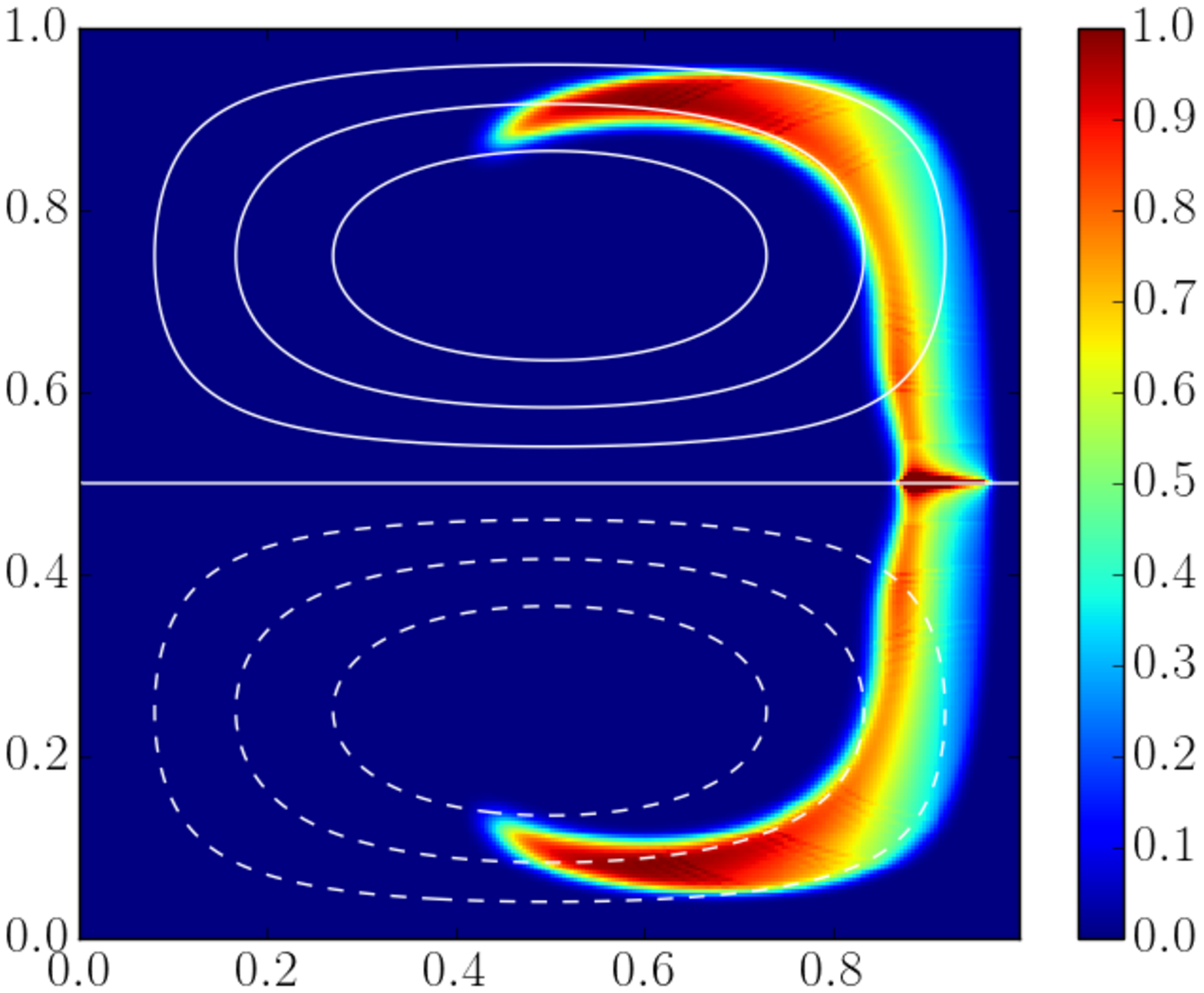}
	\ (c)\includegraphics[height=\theight, trim= 45 25 130 35, clip=true] {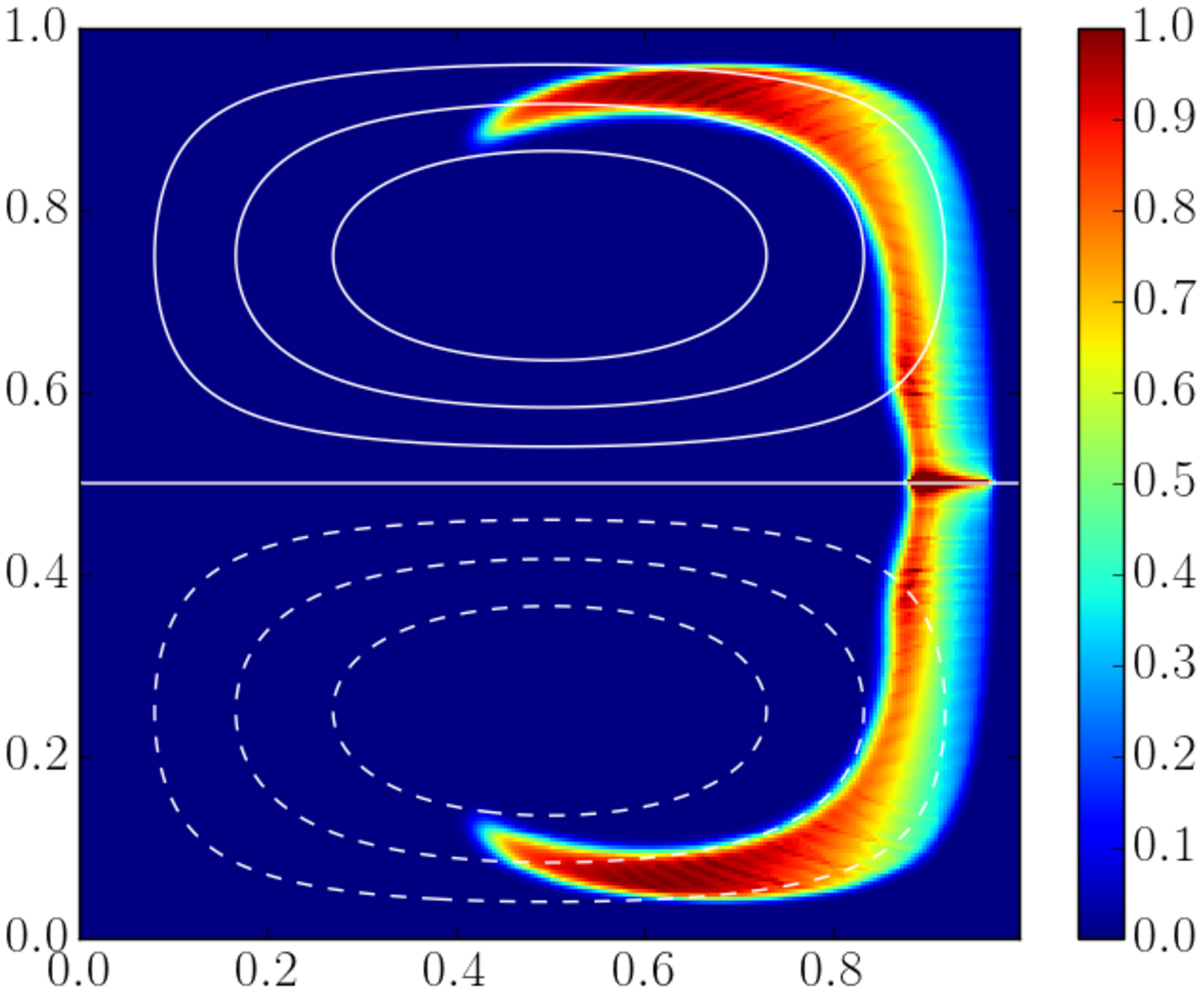}
	}
	\caption{ Comparison the profile of simulations using the first order \CCIR{} scheme 
	at a resolution of $256^2$ for different CFL: $(a):1.6 \, , (b):4.0 \, , (c):8.0$.}
\label{fig:2dRes256}
\end{figure}

At a resolution of $256^2$, the \CCIR{} is able to got accurately the profile for 
CFL number above unity. Fig.~\ref{fig:2dRes256} shows the profile computed 
for CFL up to $8$. Comparing the profiles on figs.~\ref{fig:2dCFL16e-1} and 
\ref{fig:2dRes256} with the profile of the $1024^2$ resolution simulation
at CFL$=0.8$ of fig.~\ref{fig:2dCONV}(b), the conservative schemes
are able to model the flow for CFL$>1$ with great accuracy. The excellent 
agreement between the simulation is not restricted to the profile, it also extent to
the total mass which is conserved up to machine precision. 

Semi-Lagrangian algorithms are composed of two main steps
\cite{pudykiewicz_properties_1984,smolarkiewicz_class_1992,crouseilles_conservative_2010}: 
the computation of the characteristic curves, and the reconstruction step. 
The present work was centered on making the reconstruction step conservative. 
All the simulation carried out used the $2D$-generalization of eq.~\eqref{eq:modulo}. 
Even though the trajectories were computed with a law order method, 
the algorithm can be adapted to more sophisticated methods. To do so, 
the trajectory in each point can be reconstructed using high order characteristics 
(\textit{e.g.} \cite{crouseilles_conservative_2010}) and the resulting displacement 
should be decomposed as the sum of : 
(i) a vector whose components are equal to an integer number of grid-steps, 
and (ii) a remainder vector whose components are smaller than the grid-step.

\section{Perspectives}
We have introduced a systematic approach to derive a conservative
scheme without the need for a finite volume discretisation. The method
has been successfully applied to semi-Lagrangian schemes, which are
notorious for being very efficient, but usually not conservative.
Using this method, we were able to built a third order conservative 
semi-Lagrangian scheme based on the scheme introduced by
Dahlquist and Björck.

The approach presented here is similar in the spirit to that introduced by 
Verstappen {\it et al.} in \cite{verstappen_symmetry-preserving_2003} to 
derive energy preserving schemes. They also used an adjoint formulation 
to derive the discrete scheme. As their concern is the conservation
of energy, they insist of the skew symmetry property of the operator. 
We are here rather concerned with mass conservation and therefore 
focus on the column-balance property of the scheme.

Our approach also bears similarities with ideas introduced by Shashkov in 
the support operators method \cite{shashkov_conservative_1995} or by 
Carpenter \cite{carpenter_stable_1999,fisher_discretely_2013}. 
It however differs from the support operator 
method, in that we propose an algorithm (via the discretisation of the adjoint 
equation) to systematically transform a non-conservative advection scheme 
into a continuity preserving operator.

\appendix
\section{Convergence study}
\begin{figure}
	\centerline{\subfigure[Growth-rate]{
	\includegraphics[width=0.4\textwidth, trim= 2 2 45 30, clip=true] {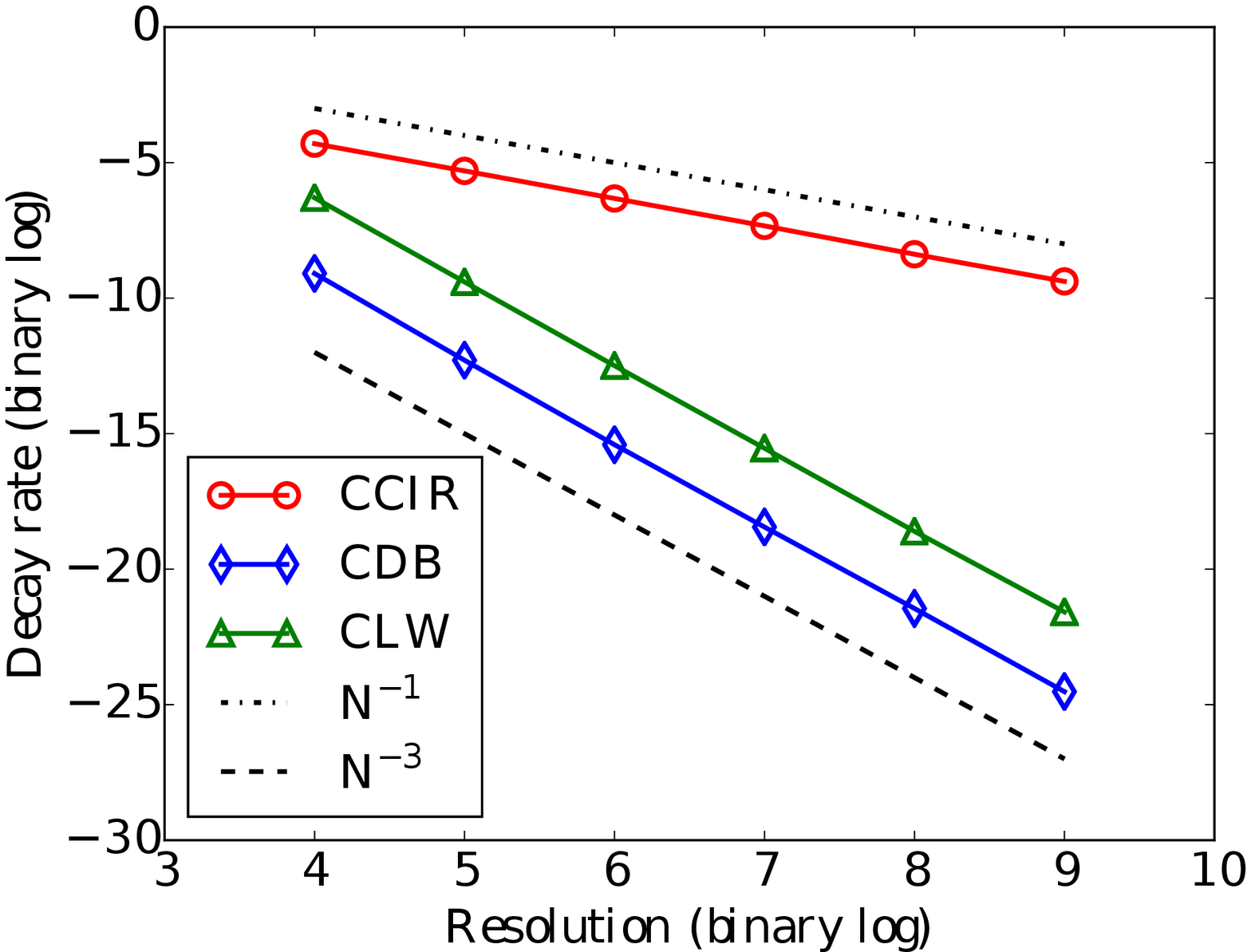}
	}\subfigure[Phase-drift]{
	\includegraphics[width=0.4\textwidth, trim= 2 2 45 30, clip=true] {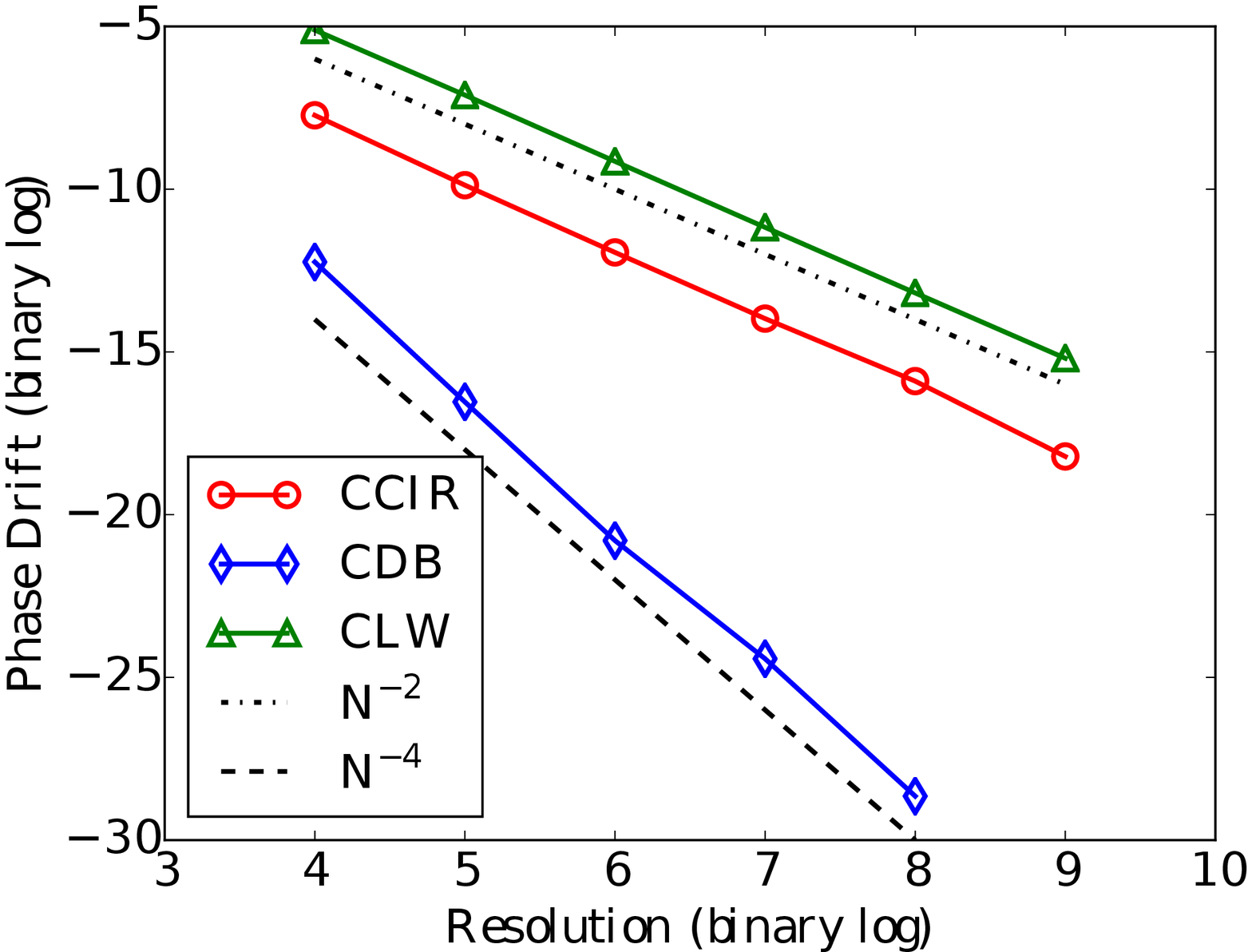}
	}} \caption{Decay rate (loss in amplitude) and phase shift 
	per unit of time for a 2D
	test case, the initial distribution takes the form $\cos(x+y)$ 
	and the flow is uniform with $u_x=u_y=1$.}
\label{fig:appx}
\end{figure}

In order to illustrate the order of convergence of the conservative schemes 
introduced in section~\ref{sec:consSL} in $2$D, we perform a numerical study 
with varying resolution. The initial distribution takes the form $\cos(x+y)$ 
and the flow is uniform with $u_x=u_y=1$. The results are illustrated in 
fig.~\ref{fig:appx} and in tables~\ref{table:appx1} and \ref{table:appx2}.

\begin{table}
\begin{center}
\begin{tabular} {c | c | c | c}
  Resolution & CCIR & CLW &  CDB \\
  \hline
     16 & 0.050728 & 0.012718 & 0.0018352 \\ 
     32 & 0.025286 & 0.0014777 & 0.00019944 \\ 
     64 & 0.012492 & 0.00017356 & 2.3e-05 \\ 
    128 & 0.0061982 & 2.1e-05 & 2.8e-06 \\ 
    256 & 0.0030018 & 2.51e-06 & 3.49e-07 \\ 
    512 & 0.0014981 & 3.19e-07 & 4.16e-08 \\ 
\end{tabular} 
\end{center}
\caption{Decay rate for the 2D advection test.}
\label{table:appx1}
\end{table}
\begin{table}
\begin{center}
\begin{tabular} {c | c | c | c }
  Resolution & CCIR & CLW & CDB \\
  \hline 
     16 & 0.0047106 & 0.028943 & 0.00020718 \\ 
     32 & 0.0010636 & 0.0072473 & 1.0516e-05 \\ 
     64 & 0.00025402 & 0.0017658 & 5.4912e-07 \\ 
    128 & 6.2017e-05 & 0.00043328 & 4.4307e-08 \\ 
    256 & 1.6382e-05 & 0.0001072 & 2.3758e-09 \\ 
\end{tabular} 
\end{center}
\caption{Phase shift for the 2D advection test.}
\label{table:appx2}
\end{table}

\bibliographystyle{siam}
\nocite{*}
\bibliography{slcs}

\end{document}